\documentclass[11pt]{article}

\def\diag{{\rm diag}}
\def\pconv{\smash{\mathop{\longrightarrow}\limits^p}}     %Converges in prob
\def\dconv{\smash{\mathop{\longrightarrow}\limits^d}}     %Converges            %in dist.

\def\var{\mbox{var}}

\def\diag{\mbox{diag}}

\def\argmin{\mbox{argmin}}
\renewcommand{\bar}{\overline}
\newcommand{\norm}[1]{\left\Vert#1\right\Vert}

\renewcommand{\tilde}{\widetilde}
\renewcommand{\hat}{\widehat}

\def\tr{\mbox{trace\;}}

\def\plim{\mbox{plim }}

\newcommand{\beq}{\begin{eqnarray*}}
\newcommand{\eeq}{\end{eqnarray*}}

\def\1T{frac{1}{T}}
\def\1n{\frac{1}{n}}

\def\bitem{\medskip\begin{itemize} \itemsep=8.0pt \parskip=8.0pt}
\def\eitem{\end{itemize}}

\usepackage{fullpage,amssymb,amsmath,mathtools,natbib,xcolor,soul}
\usepackage{natbib}

\newtheorem{lemma}{Lemma}
\newtheorem{proposition}{Proposition}

\def\sparse{S}
\def\low{L}
\def\soft{D}
\def\data{Z}
\def\noise{W}

\def\tr{\text{trace}}
\def\Fs{ F^*}
\def\Lambdas{\Lambda^*}
\def\es{ e^*}
\def\Hapca{\tilde{ H}}
\def\Fapca{\tilde{ F}}
\def\Lapca{\tilde{\Lambda}}

\def\Fpca{\hat{ F}}
\def\Lpca{\hat{\Lambda}}
\def\Dpca{D_r}
\def\Frpca{\bar F}
\def\Lrpca{\bar \Lambda}
\def\Hpca{\hat H}
\def\Lcpca{\bar\Lambda_{\gamma,\infty}}
\def\Lridge{\bar\Lambda_{\gamma,0}}

\def\pconv{\smash{\mathop{\longrightarrow}\limits^p}}     %Converges in prob
\def\pr{{^\prime }}
\def\Lp{\Lambda^{0\pr}}
\def\Fp{F^{0\pr}}
\def\vec{\mathrm{vec}}
\begin{document}

%\title{{\bf{\normalsize{REGULARIZED ESTIMATION AND INFERENCE OF APPROXIMATE FACTOR MODELS}}}}

\title{{\bf{\normalsize{PRINCIPAL COMPONENTS AND REGULARIZED ESTIMATION OF FACTOR MODELS }}}}

\author{Jushan Bai\thanks{Columbia University, 420 W. 118 St. MC 3308, New York, NY 10027.
Email: jb3064@columbia.edu}
  \and Serena Ng\thanks{Columbia University and NBER, 420 W. 118 St. MC 3308,
  New York, NY 10027. Email: serena.ng@columbia.edu
\newline
This paper was presented at the Big Data in Predictive Dynamic Econometric Modeling at UPenn, Columbia University, and at
Harvard/MIT. We thank the seminar participants and Zongming Ma for
for helpful comments. This work is supported by the National Science
Foundation SES-1658770 (Bai), SES-1558623 (Ng).}}
\date{\today\bigskip}
%\date{December 2013 \\ \bigskip  Comments Welcome}
\maketitle

\begin{abstract}
It is known that the common factors in a large panel of data can be consistently estimated by the method of principal components, and principal components can be constructed by iterative least squares regressions.  Replacing least squares with ridge regressions turns out to have the effect of shrinking the singular values of the common component and possibly reducing its rank. The method is used in the machine learning literature to recover low-rank matrices. We study the procedure from the perspective of estimating a minimum-rank approximate factor model. We show that the constrained factor estimates are biased but can be more efficient in terms of mean-squared errors.  Rank consideration suggests a data-dependent penalty for selecting the number of factors. The new criterion is more conservative in cases when the nominal number of factors is inflated by the presence of weak factors or large measurement noise. The framework is extended to incorporate a priori linear constraints on the loadings.  We provide asymptotic results that can be used to test economic hypotheses.
\end{abstract}
\bigskip


\thispagestyle{empty}
\setcounter{page}{0}
\baselineskip=18.0pt
\bibliographystyle{dcu}

\newpage
\section{Introduction}
A low rank component is characteristic of many economic data. In analysis of  international business cycles, the component arises because of global shocks. In portfolio analysis, the component arises because of non-diversifiable risk. One way of modeling this component when given a panel of data $X$ collected for $N$ cross-section units over a span of $T$ periods is to impose a factor structure. If the data have $r$ factors, the $r$ largest population eigenvalues of $XX'$ should increase with $N$. In a big data (large $N$ large $T$) setting, it has been shown that the space spanned by the factors can be consistently estimated by the eigenvectors corresponding to the $r$ largest eigenvalues of $XX^\prime$ or $X\pr X$.  But it is not always easy to decisively separate the small from the large eigenvalues from the data. Furthermore, the eigen-space is known to be sensitive to outliers even if they occur infrequently. Sparse spikes, not uncommon in economic and financial data,  may inflate the estimated number of factors.   It would be useful to recognize  such variations in factor estimation.

It is known that eigenvectors, and hence the factor estimates, can be obtained by iterative least squares regressions of $X$ on guesses of the factor scores and of the loadings.
 This paper considers an estimator that can be understood as performing iterative ridge  instead of least squares regressions. Ridge regressions are known to shrink the parameter estimates of a linear model towards zero. They are biased but are less variable. In the present context,   iterative ridge regressions will shrink the singular values of the common component towards zero.  \citet{hmlz} shows that when combined with a cleanup step that explicitly sets the small singular values to zero,  it implements {\em singular value thresholding} (SVT) and  delivers robust principal components (RPC) as output.
 %SVT is an operation developed within the last decade to solve the surrogate of NP hard problems that entail rank minimization.

 Our interest in SVT stems from its ability to estimate approximate factor models of minimum rank. Researchers have long been interested in minimum-rank factor analysis, though the effort to find a solution has by and large stopped in the 1980s because of computationally challenges. SVT overcomes the challenge by solving a relaxed surrogate problem and delivers a  robust estimate of the  common component. But while worse case error bounds for RPC  that are uniformly valid over models in that class are available, these algorithmic properties make no reference to the  probabilistic structure of the data. Their use in classical statistical inference is limited.   We approach the problem from the perspective of a parametric factor model. Since we make explicit assumptions about the data generating process, we can obtain parametric rates of convergence and make precise the  effects of singular value thresholding on the factor estimates. Our results are asymptotic, and present an alternative perspective to the algorithmic ones obtained under the assumption of a fixed sample.

 This paper makes several contributions. The first is to provide a statistical analysis of RPC that complements  the results developed from a machine learning perspective.\footnote{The literature on PC is vast. See, for example, \citet{jolliffe-book}. Some recent papers on SVT are \citet{uhzb}, \cite{agarwal-negahban-wainwright:12}, \citet{yang-ma-buja:14},  \citet{hmlz}.} Constrained estimation generally leads to estimates that are less variable, but at the cost of bias. As we will see, rank constrained estimation is no exception.  Our second contribution is to provide a frequentist framework for regularized factor analysis. Economic theory often suggests a priori restrictions in the form of single or cross-equation restrictions.  We  provide the inferential theory that permits testing of economic hypothesis in the form of general linear restrictions on the loadings, with or without rank constraints.

 Our third contribution is to incorporate minimum rank consideration into the selection of the number of factors.  We propose a new criterion that implicitly adds a data dependent term due to the desire for a minimal rank common component to the deterministic penalty introduced in \citet{bai-ng-ecta-02}.  Simulations suggest that the resulting criterion gives a more conservative estimate of the number of factors when there are outliers in the data, and when the contributions of some factors to the common component are small. An appeal of the new procedure is that we do not need to know which the assumptions in the factor model are violated.

 The follow notation will be used in what follows. We use the  $(T,N)$ to denote sample size  of $X$ in statistical factor analysis,  but  $(m,n)$ to denote dimension of a matrix $Z$ when we are considering algorithms. For an arbitrary $m\times n$ matrix $Z$, the full singular value decomposition (\textsc{svd}) of $Z$ is  $ Z=U D V\pr$ where $U=[u_1,\ldots, u_m)$ is $m\times m$ and $V=[v_1,\ldots, v_n]$ is $n\times n$, $U\pr U=I_m$, $V\pr V=I_n$, and $D$ is a $m\times n$ matrix of zeros except for its $\min(m,n)$ diagonal entries which are taken by the non-zero population singular values $d_1,d_2,\ldots, d_{\min(m,n)}$ of $Z$. The left eigenvectors of  $ZZ\pr$ are the same as the left singular vectors  of $Z$ since $ZZ\pr= U D^2 U\pr$. The nuclear norm $\norm{Z}_*=\sum_{i=1}^{n}d_i(Z)$ is  the sum of the singular values of $Z$. The singular values are ordered such that $d_1(Z)$ is the largest.  Let $\norm{Z}_1=\sum_{i,j}|Z_{ij}|$ be the component-wise 1-norm, and let $\norm{Z}^2_F=\sum_{i=1}^m \sum_{j=1}^n |Z_{ij}|^2$ denote the  Frobenius (or component-wise-2) norm.   Let  $U=[U_r; U_{n-r}]$ and $V=[V_r, V_{n-r}]$ where $U_r$ consists of the first $r$ columns, while  $U_{n-r}$ consists of the last $(n-r)$ columns of $U$.  A similar partition holds for $V$. Then $ Z=U_r D_r V_r\pr + U_{n-r}D_{n-r} V_{n-r}\pr$.
 % and the truncated SVD of $Z$, being $U_rD_r V_r\pr$ if $d_j=0$ for $j>r$. If the rank of $Z$ is no smaller than $r$, then
 %$U_r D_r V_r'$ is the optimal rank-$r$ approximation of $Z$ under the Frobenius norm.

 An important step in our analysis is to make use of results derived previously for the method asymptotic principal components (APC). In the next section, we show that  variants of principal components  differ from APC by the normalization used.

\section{Estimation of Approximate Factor Models}
\label{sec:sec2}

We use  $i=1,\ldots N$ to index cross-section units and $t=1,\ldots T$ to index time series observations. Let  $X_i=(X_{i1},\ldots X_{iT})^\prime$ be a $T\times 1$ vector of random variables.
%standardized data. Hence each $X_i$ is mean zero and have unit variance.
The $T\times N$ data matrix  is denoted $X=(X_1,X_2,\ldots, X_N)$. The  factor representation of the data is
    \begin{eqnarray}
    \label{eq:dgp}
    X&=&F^0\Lambda^{0^\prime} + e
    \end{eqnarray}
where   $F$ is a $T\times r$ matrix of common factors,  $\Lambda$ is a $N \times r$ matrix of factor loadings and whose true values are $F^0$ and $\Lambda^0$. We observe $X$, but not $F$, $\Lambda$, or $e$. The variations in the common component $C=F\Lambda^\prime$ are pervasive, while those in  the idiosyncratic errors $e$ are specific.  The population covariance structure of $X_t=(X_{1t},X_{2t},...,X_{Nt})'$ is
\[ \Sigma_X=\Sigma_C+\Sigma_e.
\]
where $\Sigma_C=\Lambda \Sigma_F \Lambda'$.
A strict factor model assumes that $\Sigma_e$ is diagonal.  Under the assumption that $T$ tends to infinity with $N$ fixed,  \citet{anderson-rubin}, \citet{joreskog:67}, and \citet{lawley-maxwell} show that the  factor loadings estimated by maximum likelihood or covariance structure methods  are $\sqrt{T}$ consistent and asymptotically normal.

\subsection{Asymptotic Principal Components (APC): $(\Fapca,\Lapca)$}
The assumption that $\Sigma_e$ is diagonal is restrictive for many economic applications. The approximate factor model of \citet{chamberlain-rothschild} relaxes this assumption. A defining characteristic of an approximate factor  model with $r$ factors is that  the $r$ largest population eigenvalues of $\Sigma_X$ diverges as $N$ increases, while the $r+1$ largest eigenvalue is bounded.
We study estimation of an approximate factor model under the assumptions in \citet{bai-ng-ecta-02} and \citet{bai-ecta-03}.
\paragraph{Assumption A} There exists a constant $M<\infty$ not depending on $N,T$ such that
\begin{itemize}
	\item[a.] (Factors and Loadings): $E||F_t^0||^4 \le M$, $||\Lambda_i||\le \bar \Lambda$,
$\frac{F^{0\pr}F^0}{T}\pconv \Sigma_F >0$, and $ \frac{\Lambda^{0\pr}\Lambda^0 }{N}\pconv \Sigma_{\Lambda}>0$.

\item[b.] (Idiosyncratic Errors): Time and cross-section dependence
\begin{itemize}
\item[(i)] $E(e_{it})=0, E|e_{it}|^8\le M$;
\item[(ii)] $E(\frac{1}{N}\sum_{i=1}^N e_{it} e_{is})=\gamma_N(s,t)$, $|\gamma_N(s,s)|\le M$ for all $s$ and $\frac{1}{T}\sum_{s=1}^T \sum_{t=1}^T |\gamma_N(s,t)|\le M$;
\item[(iii)] $E(e_{it}e_{jt})=\tau_{ij,t}$, $|\tau_{ij,t}|\le | \tau_{ij,t}|$ for some $ \tau_{ij,t}$ and for all $t$, and $\frac{1}{N}\sum_{i=1}^N \sum_{j=1}^N |\tau_{ij,t}|\le M$;
\item[(iv)] $E(e_{it}e_{js})=\tau_{ij,st}$ and $\frac{1}{NT}\sum_{i=1}^N \sum_{j=1}^N \sum_{t=1}^T \sum_{s=1}^T |\tau_{ij,ts}|<M$;
\item[(v)]  $E|N^{-1/2}\sum_{i=1}^N |\sum_{i=1}^N [e_{is}e_{it}-E(e_{is}e_{it})]^4\le M$ for every $(t,s)$.
\end{itemize}
\item[c.] (Central Limit Theorems): for each $i$ and $t$,
$	\frac{1}{\sqrt{N}} \sum_{i=1}^N \Lambda^0 _i e_{it}\dconv N(0,\Gamma_t)$ as  $N\rightarrow\infty$, and
$	\frac{1}{\sqrt{T}}\sum_{t=1}^T F^0_te_{it}\dconv N(0,\Phi_i) $ as  $T\rightarrow \infty$.
\end{itemize}

Assumption A allows the factors  to be dynamic and the errors to be serially and cross-sectionally dependent as well as heteroskedastic. The loadings can be fixed or random. While $\Sigma_e$ need not be a diagonal matrix, (b) also requires it to  be sufficiently sparse (the correlations to be weak). Thus Assumption A imposes a strong factor structure via positive definiteness of $\Sigma_F$ and $\Sigma_{\Lambda}$.
Part (a) and (b) imply weak dependence between the factors and the errors:  $E(\frac{1}{N}\sum_{i=1}^N ||\frac{1}{\sqrt{T}} \sum_{t=1}^T F_t^0e_{it}||^2)\le M$. \citet{bai-ng-ecta-02} shows that $r$ can be consistently estimated. In estimation of the $F$ and $\Lambda$, the number of factors $r$ is typically treated as known.

For given $r$, the method of APC solves the following  problem:
 \begin{eqnarray}
  \min_{F,\Lambda, \frac{F\pr  F}T=I_r} \frac{1}{NT}\sum_{i=1}^N \sum_{t=1}^T (X_{it}-\Lambda_i\pr F_t)^2=\min_{F,\Lambda} \frac{1}{NT}\norm{X-F\Lambda\pr}_F^2.
  \label{eq:ols-obj-apca}
  \end{eqnarray}
If we concentrate out $\Lambda$ and use the normalization $\frac{F\pr  F}T=I_r$,  the problem is the same as maximizing $\textsc{tr} (F\pr (XX\pr )F$ subject to $F'F/T=I_r$. But the solution is not unique. If $F$ is a solution, then $FQ$ is also a solution for any orthogonal $r\times r$ matrix $Q$ ($QQ'=I_r$).  However, if we put the additional restriction that $\Lambda'\Lambda$ is diagonal, then the solution becomes unique (still up to a column sign change).

The APC estimates, denoted  $(\Fapca,\Lapca)$, are defined as\footnote{The non-zero eigenvalues $XX'$  and $X' X$ are the same. An alternative estimator is based on the eigen-decomposition of the $N\times N$ matrix $X\pr X$ with normalization $\frac{\Lambda\pr\Lambda}{N}=I_r$.
%Denote $(\tilde F^T,\tilde \Lambda^\prime)$ the solutions based on $XX\pr$ and
%Let $(\tilde F^\dag,\tilde \Lambda^\dag)$ denote the solutions based on $(X\pr X)$.
%It is easy to see that $\tilde F^\dag=\tilde F \tilde D_r$ \hl{changed to $\tilde D_r$} and $\tilde \Lambda^\dag=\tilde \Lambda \tilde D_r^{-1}$ \hl{changed to $\tilde D_r$}.   The results to follow are based on the decomposition of the $T\times T$ matrix $XX\pr$, that is, we focus on $(\tilde F,\tilde \Lambda)$.
}
 \begin{subequations}
 \begin{eqnarray}
\Fapca&=&\sqrt{T}  U_r     \label{eq:F-apca}\\
   \Lapca&=&X\pr \Fapca/T    \label{eq:L-apca}
\end{eqnarray}
\end{subequations}
That is,  the matrix of factor estimates is $\sqrt{T}$ times the eigenvectors corresponding to  the $r$ largest eigenvalues of $XX\pr/(NT)$.
%Let $D_r^2=\diag(d_1^2, d_2^2,...,d_r^2)$ with $d_1^2>d_2^2>\cdots d_r^2$ being the first $r$ largest eigenvalues of $XX'/(NT)$.
It can be verified that $\tilde \Lambda'\tilde \Lambda/N= D_r^2$, a diagonal matrix of the $r$ largest eigenvalues of $\frac{XX^\prime}{NT}$.  \citet{bai-ng-ecta-02} shows that    as  $N,T\rightarrow \infty$,
\[\min(N,T)\frac{1}{T}\sum_{t=1}^T ||\Fapca_t -\Hapca_{NT} F^0_t||=O_p(1)  .\]
That is to say,  $\Fapca_t$ consistently estimates  $F^0_t$ up to a rotation  by the matrix $\Hapca_{NT}$,
 defined  as
 \begin{equation}\label{tildeH_NT} \Hapca_{NT}=
 \bigg(\frac{\Lambda^{0\pr}\Lambda^0}{N}\bigg)\bigg(\frac{ F^{^0 \pr} \Fapca}{T}\bigg)\Dpca^{-2}\end{equation}

This `big data blessing' has generated a good deal of econometric research in the last two decades.\footnote{The method of APC is due to \citet{connor-korajczyk-86}. \citet{fhlr-restat} and \citet{stock-watson-di,stock-watson-diforc} initiated interests in large dimensional factor models. See \citet{bai-ng-survey} for a review of this work. \citet{fan-liao-mincheva} shows consistency of the factor estimates when the principal components are constructed from the population covariance matrix.}  As
as explained in \citet{bai-ng:joe-13},  $\Hapca_{NT}$ will not, in general, be an identity matrix implying that the $j$-th factor estimate $\Fapca_j$ will not, in general, equal $F^0_j$ even asymptotically. The exception is when the true $F^0$ is such that $\frac{F^{0\pr}F^0}{T}=I_r$ and $\Lambda^{0'}\Lambda^0$ is diagonal. Of course, it would be unusual for $F^0$ to have second moments that agree with the normalization used to obtain $\Fapca$. Nonetheless, in applications when interpretation of $F$ is not needed as in forecasting, the fact that $\Fapca$ consistently estimates the space spanned by $F^0$  enables  $\Fapca$ to be used as though $F^0$ were observed.

 Theorem 1 of  \citet{bai-ecta-03}  shows that if $\sqrt{N}/T\rightarrow 0$
as $ N,T\rightarrow \infty$, then  $\plim_{N,T\rightarrow\infty} \frac{\Fapca\pr F^0}{T}=\mathbb Q_r,$
$\plim_{N,T\rightarrow\infty} D_r^2 = \mathbb D_r^2$, and
\begin{eqnarray*}
\sqrt{N} (\Fapca_t- \Hapca^\prime_{NT} F_t^0)&
\dconv&  \mathcal N\bigg(0,   \mathbb D_{r}^{-2}\mathbb Q_r \Gamma_t \mathbb Q_r\pr \mathbb D_{r}^{-2}\bigg)
\equiv \mathcal N(0,\textsc{Avar}(\Fapca_t))\\
\sqrt{T} (\Lapca_i- \Hapca_{NT}^{-1} \Lambda_i^0)&=
\dconv & \mathcal N\bigg(0,  (\mathbb Q_r\pr)^{-1} \Phi_i \mathbb Q_r^{-1} \bigg)\equiv \mathcal N(0,\textsc{Avar}(\Lapca_i))
\end{eqnarray*}
where $\mathbb Q_r=\mathbb D_{r} \mathbb V_{r}\Sigma_{\Lambda}^{-1/2}$, and
 $\mathbb D_{r}^2$ and $ \mathbb V_{r } $ are the eigenvalues and eigenvectors of the $r\times r$ matrix $\Sigma_{\Lambda}^{1/2}\Sigma_F\Sigma_{\Lambda}^{1/2}$, respectively.
 %\textcolor{red}{ This $\Sigma_{\mathbb C}$ resembles the covariance of the true component component. All I want to understand here is its  relation to  $\lim_{N,T\rightarrow \infty}\Sigma_{\hat C}$. If $\tilde H_\infty=I_r$, will they ever be identical ? There is no need for a Lemma if you don't want it.}
The asymptotic inference of the factors ultimately depends on the eigenvalues and eigenvectors of the true common component. Nonetheless, as shown in \citet{bai-ng-ecta-06},  $\Fapca$ can be used in subsequent regressions  as though they were $F^0$ provided $\frac{\sqrt{T}}{N}\rightarrow 0$.

While $\tilde H_{NT}$ is widely used  in asymptotic analysis, it is difficult to interpret.
It is useful to consider the following asymptotically equivalent rotation matrices:
\begin{lemma}
\label{lem:HNT}
Let $ \tilde H_{1,NT}=(\Lambda^{0\prime}\Lambda^0)(\Lapca^{\prime}   \Lambda^0)^{-1}$ and $ \tilde H_{2,NT}=(F^{0\prime}F^0)^{-1}(F^{0\prime} \Fapca)$.  Then
(i): $ \tilde H_{NT}=\tilde H_{1,NT}+o_p(1)$  and (ii):
$\tilde H_{NT}=\tilde H_{2,NT}+o_p(1)$.
\end{lemma}
From Lemma \ref{lem:HNT},  we see that
the inverse of $\tilde H_{1,NT}$ is the regression coefficient  of
$\tilde \Lambda$ on $\Lambda^0$, while $\tilde H_{2,NT}$ is  the regression coefficient of $\tilde F$ on $F^0$.
The $o_p(1)$ term in the lemma can be shown to be $O_p(1/\min(N,T))$.
Theorem 1 of \citet{bai-ecta-03} remains valid when $\tilde H_{NT}$ is replaced by either $\tilde H_{1,NT}$ or $\tilde H_{2,NT}$.  In addition to being interpretable,
these simpler rotation matrices may simplify proofs in future work, hence of independent interest.

\subsection{Principal Components (PC): ($\Fpca,\Lpca)$}
Whereas the eigenvectors of $XX'$ are known in the economics literature as APC,  principal components (PC) are sometimes associated with the singular vectors of $X$. In statistical modeling, APC tends to emerge from a  spiked-covariance analysis, while PC tends to follow from a spiked-mean analysis.  At a more mechanical level,  $\tilde F$ defined above depends only on the eigenvectors but does not depend on $D_r$. This is a somewhat unusual definition, as textbooks such as \citet{hastie-tibs-friedman} define principal components as $U_rD_r$. Nonetheless, both definitions are valid and differ in the normalization used. We now consider yet another definition of principal components for reasons that will soon be obvious.

If we write the SVD of $X$ as $X =U\breve{D}V'$,
 the singular values in $\breve{D_r}$ are of $O_p(\sqrt{NT})$ magnitude. To facilitate asymptotic analysis,  we consider the scaled data
 \[Z=\frac{X}{\sqrt{NT}}, \quad\quad \textsc{svd}(Z)=U D V', \quad D=\frac{\breve D}{\sqrt{NT}}.\]
Note that the left and right singular vectors of $Z$ are the same as those for $X$.  However,  while the  first $r$ singular values of $X$ (i.e. $\breve D_r$) diverge and the
 remaining $N-r$ are bounded as $N,T\rightarrow \infty$,  the $r$ singular values
 of $Z$ (i.e. $D_r$) are bounded and the remaining $N-r$ singular values tend to zero.

The  model for the scaled data $Z$ is
%     \begin{eqnarray*}
%    \Fupca= U_r\breve D_r^{1/2}, \quad\quad
%    \Lupca\pr=  \breve D_r^{1/2} V_r\pr.
%    \end{eqnarray*}
\begin{equation}
 Z = \Fs {\Lambdas}^\prime+\es
\label{eq:scaled-model}
\end{equation}
where
$\Fs=\frac{F^0}{\sqrt{T}}$, $ \Lambdas=\frac{\Lambda^0}{\sqrt{N}}$, and $\es=\frac{e}{\sqrt{NT}}$.
Based on the  \textsc{svd} of $Z=UDV\pr$, we  define the PC estimates as:
\begin{subequations}
	\begin{eqnarray}
	\Fpca_z&=&  U_r \Dpca^{1/2}\label{eq:F-pca-Z}\\
	\Lpca_z&=& V_r  \Dpca^{1/2}  \label{eq:L-pca-Z}.
	\end{eqnarray}
\end{subequations}

Notably, the PC and the  APC estimates are equivalent up to a scale transformation. In particular,
$ \Fpca_z=
 \Fapca \frac{\Dpca ^{1/2}}{\sqrt{T}} $ and
$ \Lpca_z=   \Lapca \frac{\Dpca ^{-1/2}}{\sqrt{N}}$.
One can construct the APC factor estimates directly from a \textsc{svd} of
$XX\pr$  and rescale the eigenvalues, or one can construct the  PC factor estimates from a  $\textsc{svd}$ of the rescaled data $Z$.\footnote{There may be numerical advantages to using PC over APC. The documentation of \textsc{prcomp} in \textsc{R} notes 'the calculation is done by a singular value decomposition of the (centered and possibly scaled) data matrix, not by using eigenvalues on the covariance matrix. This is generally the preferred method for numerical accuracy.}  While $(\Fpca_z,\Lpca_z)$ emerges from the optimization problem of
$\min_{F,\Lambda}\|Z-F\Lambda'\|_F^2$,
$\Fpca_z$ is an estimate for $F^*=\frac{F^0}{\sqrt{T}}$, not for $F^0$. Similarly, $\Lpca_z$ is an estimate for $\Lambda^*=\frac{\Lambda^0}{\sqrt{N}}$. For estimation of $F^0$ and $\Lambda^0$, we define
\begin{subequations}
	\begin{eqnarray}
	\Fpca&=\sqrt{T} \, U_r \Dpca^{1/2}\label{eq:F-pca}\\
	\Lpca&=\sqrt{N}\, V_r  \Dpca^{1/2}  \label{eq:L-pca}.
	\end{eqnarray}
\end{subequations}
That is to say, $\Fpca=\sqrt{T}\hat F_z$ and $\Lpca=\sqrt{N}\hat \Lambda_z$. It follows that
% The normalized estimator is the solution of
%$\min_{F,\Lambda}\|X-F\Lambda\|_F^2$, which is the same as Lemma 1 with $Z$ replaced by $X$.
%\footnote{Note that we can simply define $\Fpca=U_r \Dpca^{1/2}$ and $\Lpca=V_r \Dpca^{1/2}$ so that
%$\Fpca'\Fpca =\Lpca'\Lpca=D_r$, but this $\Fpca$ estimates $F^*$ instead of $F^0$.}
\[ \frac{\Fpca'\Fpca} T =\frac{ \Lpca'\Lpca} N = D_r, \quad\quad \hat F = \tilde F \Dpca^{1/2}, \quad \quad \hat \Lambda = \tilde \Lambda \Dpca^{-1/2}  \]
 In contrast,  APC uses the normalization $\frac{F\pr F}{T}=I_r$,  and $\Lambda\pr\Lambda $ is diagonal. The   unit length normalization makes it inconvenient to impose restrictions on the APC estimates of $F$, a limitation that is important for the analysis to follow.

To establish the large sample properties of the PC estimates, we need to compare the estimates with a rotation of the true factors. Given the relation between $\Fapca $ and $\Fpca$, it is natural to define the new rotation matrix as
\[ \hat H_{NT}= \tilde H_{NT} \Dpca^{1/2}. \]
This leads to the identities
\begin{eqnarray*}
 \sqrt{N} (\Fpca_t - \Hpca^\prime _{NT} F_t^0)&=&
 \sqrt{N} \Dpca^{1/2}  (\Fapca_t - \Hapca_{NT}\pr F_t^0),\\
 \sqrt{T} (\Lpca_i - \Hpca_{NT}^{-1} \Lambda_i^0)&=&
 \sqrt{T} \Dpca^{-1/2}  (\Lapca_i - \Hapca_{NT}^{-1} \Lambda_i^0).
\end{eqnarray*}
From the limiting distributions of $\tilde F_t$ and $\tilde \Lambda_i$, we obtain:
\begin{proposition}
\label{prop:prop1}
Suppose that the data are generated by (\ref{eq:dgp}) and Assumption A holds and
both $N$ and $T$ go to infinity. Then  the PC-estimates $\Fpca$ and $\Lpca$ satisfy:
\begin{itemize}
\item[(i)] $\sqrt{N}(\Fpca_t-\Hpca^\prime_{NT} F^0_t)\dconv N\bigg(0,\, \mathbb D_{r}^{1/2}\,
\textsc{Avar}(\Fapca_t) \, \mathbb D_{r}^{1/2}\bigg) \equiv N\bigg(0,\textsc{Avar}(\Fpca_t)\bigg)$;
\item[(ii)] $\sqrt{T}(\Lpca_i -\Hpca_{NT}^{-1}\Lambda^0_i) \dconv N\bigg(0, \, \mathbb D_{r}^{-1/2} \, \textsc{Avar}(\Lapca_i)\, \mathbb D_{r}^{-1/2} \bigg)\equiv N\bigg(0,\textsc{Avar}(\Lpca_i)\bigg)$;
\item[(iii)] Let
$A_{NT} (\hat C_{it})= \frac 1 T \hat\Lambda_i' \textsc{Avar}(\hat F_t) \hat \Lambda_i + \frac 1 N \hat F_t' \textsc{Avar}(\hat \Lambda_i) \hat F_t.$  Then
$ (A_{NT} (\hat C_{it})) ^{-1/2}(\hat C_{it}-C_{it}^0 ) \dconv N(0,1) $.

\end{itemize}
\end{proposition}
The PC estimate of $F_t$ and $\Lambda_i$ are $\sqrt{N}$ and $\sqrt{T}$ consistent respectively, just like the APC estimates. It follows that the estimated common component $\hat C_{it}$ is $\min(\sqrt{N},\sqrt{T})$ consistent.
Analogous to Lemma 1, we can define asymptotically equivalent rotation matrices. Let
$\hat H_{1,NT}=(\Lambda^{0\prime}\Lambda^0)(\Lpca^{\prime}   \Lambda^0)^{-1}$
and $\hat H_{2,NT}=(F^{0\prime}F^0)^{-1}(F^{0\prime} \hat F)$. Proposition \ref{prop:prop1} holds
with $\hat H_{NT}$ replaced by either $\hat H_{1,NT}$ or $\hat H_{2,NT}$.

\section{Rank and Nuclear-Norm Minimization}
\label{sec:sec3}

In large dimensional factor analysis, it is customary to choose an $ r\in[0,\text{rmax}]$ that  provides a good fit while taking model complexity into account.  \citet{bai-ng-ecta-02} suggests a class of consistent factor selection criteria, that is, rules that yield $\hat r$ such that $\text{prob}(\hat r=r)\rightarrow 1$ as $N,T\rightarrow \infty$. While criteria in this class have little difficulty singling out the dominant factors, they tend to be too liberal. Over-estimating the number of factors is possible when validity of Assumption A is questionable. The two leading causes are weak loadings, and idiosyncratic errors with large variances. The first has the implication that the smaller eigenvalues do not increase sufficiently fast with $N$, an issue emphasized in \citet{onatski-11} and others. The second has the implication that  $r+1$-th eigenvalue (which should not increase with $N$) is not well separated from the $r$-th eigenvalue (which should increase with $N$).  This is because outliers can increase the variance in an otherwise uninformative direction, and PCA is blind to the source of the variance, \citet{hubert-rousseeuw}. The problem is well known, but is not given attention in the econometrics literature. Both weak  loadings and  outliers can distort the number of factors being estimated. A  more conservative criteria that guard against these distortions without pre-specifying the source of over-estimation is desirable.

Our approach is based on the notion of minimum rank. A variety of problems have minimum rank as motivation.  Recall that the rank  of an arbitrary $m\times n$ matrix $Z $ is the largest number of columns of $Z$ that are linearly independent.  A related concept is the {\em spark},  which is the smallest $k$ such that $k$ columns of $Z$ are linearly independent.  The spark of a matrix is a lesser known concept because its  evaluation is  combinatorially hard. In contrast, the rank of a matrix can  be computed as the number of non-zero singular values. Nonetheless, \textsc{spark}$(Z)\le$\textsc{rank}$(Z)+1$. This implies that the rank of $Z$, in our case the number of factors, is not, in general, the smallest set of factors in $Z$. This smaller set is what we now seek to recover. We motivate
 with a discussion of  three minimum rank problems: minimum rank factor analysis, matrix completion, and low rank matrix decompositions.

\subsection{Minimum Rank Factor Analysis}
Factor analysis has its roots in the study of personality traits. The original goal of factor analysis  was  to find a decomposition for the $N\times N$ matrix $\Sigma_X$ into $\Sigma_X=\Sigma_C+\Sigma_e$, where $\Sigma_C$ can be factorized into $\Lambda\Lambda^\prime$, $\Lambda$ is $N\times r$,  and such that
\begin{equation}
\label{eq:minfac}
\text{(i)}. \quad \Sigma_X-\Sigma_e \succeq 0, \quad\quad\quad \text{(ii)}\quad \Sigma_e \succeq 0, \quad
\text{ (iii)}\quad \Sigma_e \quad\text{diagonal}.
 \end{equation}
Constraint (i) require that the low rank communality matrix  $\Sigma_C=\Sigma_X-\Sigma_e$ is positive semi-definite, and constraint (ii) requires that the Heywood case of negative error variances does not arise.  The smallest rank that solves this problem has come to be known as the minimum rank. But while it is not hard to compute $\text{rank}(X)$ from $\norm{D (X)}_0$, rank minimization is a NP hard problem because  the  the cardinality function  is non-convex and non-differentiable. Furthermore,  research such as by \citet{guttman:58} suggests that the number of factors in the data is rather large, casting doubt on the usefulness of the notion of a minimum rank. Interests in the problem subsided.

 New attempts were made to tackle the problem in the 1980s. Of interest to us are the two that involve minimizing the eigenvalues of the communality matrix. Define
\[f(\Sigma_e,;r_0)=\sum_{i=r_0}^n D_{ii,\Sigma_X-\Sigma_e}
.
\]
The first attempt is {\em constrained minimum trace factor analysis} (CMFTA)  which  finds $\Sigma_e$ to minimize $f(\Sigma_e;1)$ subject to (\ref{eq:minfac}). Its name arises from the fact that $f(\Sigma_e,1)$ is the sum of the eigenvalues of $\Sigma_X-\Sigma_e$, which is  the trace of $(\Sigma_X-\Sigma_e)$.  The second attempt is {\em approximate minimum rank factor analysis} (MFRA) which distinguishes between explained and unexplained common variance. In the MFRA setup, $\Sigma_C=\Sigma_C^++\Sigma_C^-$ where  $\Sigma_{C^+}$ is the common variance that will be explained, and has   $r$ non-zero eigenvalues that coincide with the largest eigenvalues of $\Sigma_C$.   Hence $\Sigma_X=\Sigma_C^++ (\Sigma_C^-+\Sigma_e).$
For fix $r$, MFRA then minimizes the unexplained common variance $\sum_{i=r+1}^n D_{ii,\Sigma_X-\Sigma_e}=f(\Sigma_e;r)$  subject to (\ref{eq:minfac}). In this context, \citet{tenberge-kiers} defines   the approximate minimum rank to be  the smallest  $r$ such that   $\min _{\Sigma_e} f(\Sigma_e,r)\le \delta$ with $\delta>0$.  The minimum rank   problem that the earlier literature has sought to solve  is a special case when $\delta=0$.\footnote{CMTFA is due to   \citet{bentler-72}. See also \citet{bentler-woodward,shapiro-82,shapiro-tenberge}. For MFRA, see \citet{tenberge-kiers}, and \citet{shapiro-tenberge}.} We will use $r^*$ to denote the approximate minimum rank.

What makes MFTA and MFRA  interesting is that  the sum of eigenvalues are  convex functions. Instead of tackling the original  problem that is NP hard, they solve  surrogate problems that can take advantage of interior point and semi-definite algorithms used in convex optimization.
% solve  problems of the form $\min_X(C,X)$ subject to
 %$A(X)=b$ and $X\succeq 0$ where $X$ and $C$ are $n\times n$ symmetric matrices,
  %$b\in  R^m$ and $A:S^n\rightarrow R^m$ is a linear map.   Here,
  %$\lambda_{\min}=\text{min}_x x\pr (\Omega-\Sigma_e )x$. Given $\psi$.   }
Even though the proposed algorithms for solving MTFA and MRFA are no longer efficient given today's know how,  convex relaxation of the rank function is an active area of research in recent years.   We now turn to this work.\footnote{The connection between factor analysis and low rank matrix decompositions was a focus of \citet{scpw} and a recent paper by \citet{bcm:16}.}

\subsection{Singular-Value Thresholding (SVT)}
Many problems of interest in the big data era are concerned with recovery of a low rank matrix from limited or noisy information. For example, compressive sensing algorithms seek to reconstruct a signal from a system of underdetermined linear equations. In face recognition analysis, the goal is to recover a background image from frames taken under different conditions. Perhaps the best known example of matrix recovery
is the  Netflix challenge. Contestants were given a small training sample of data on movie ratings by a subset of Netflix users and were asked to accurately predict ratings  of all movies in the evaluation sample.  Without any restrictions on the completed matrix, the problem is under-determined since the missing entries can take on any values. To proceed, it is assumed that the matrix $Z$ to be recovered has low rank. For the Netflix problem, the low rank assumption amounts to requiring that preferences are defined over  a small number of features (such as genres,  leading actor/actress).  The Netflix problem is then to complete $Z$ by finding a $L$ that is factorizable into two matrices (for preference and features) with smallest rank possible,   and such that $Z_{ij}=L_{ij}$ for each $(i,j)$ that is observed. But, as noted earlier, rank minimization  is NP-hard. The breakthrough is to replace rank minimization by the nuclear norm minimization. This is important because the nuclear norm, which is also the  sum of the eigenvalues,  is convex.   \citet{candes-recht} shows   $Z$ can be recovered  with high probability if the number of observed entries satisfies a lower bound, and that the unobserved entries are missing uniformly at random.
%if $p \ge Cr \max(m,n)^{6/5}\ln(n)$  and $r\le  \max(m,n)^{1/5}$.

 In exact matrix completion, the problem arises because of incomplete observations, but
the the data are perfectly measured whenever they are observed. A different matrix recovery problem arises not because of missing values,  but because the data are observed with errors, also referred to as noise corruption. Generically, a $m\times n$ matrix $\data$ can be decomposed as
 \[\data=\low+\sparse\]
 where $\low $  is a  matrix of reduced rank $r$, and $\sparse$ is a sparse noise matrix. A measure of sparsity is cardinality, which is the number of non-zero entries, written $\norm{\sparse}_0$.
The goal is to recover  $\low $ from data that are sparsely corrupted by noise.  While the Eckart-Young theorem states that $U_rD_rV_r\pr$ provides the best low rank approximation of $\data$ in the spectral norm, singular value decompositions are known to be sensitive to
large errors in practice, even if there are few of them. For example, if the data have fat tails, a small number of extreme values can account for a significant amount of variation in the data.\footnote{See \citet{dgk:81}, \citet{li-chen:85}, \citet{ma-genton:01}, \citet{hrvb:05}, among others.} Since PCA is blind to the source of large variations, large noise contamination can corrupt the low rank component being identified, as will be seen in the simulations to follow.

 To reduce noise corruption,  one may want to penalize $\sparse$ and solve the regulated problem
 \[\text{rank}(\low)+\bar \gamma \|\sparse\|_0, \quad\quad s.t. \quad \data=\low+\sparse.\]
  This is challenging because rank and cardinality are both non-convex functions. \citet{wright-ma-rao,candes-li-ma-wright} show that under an incoherence on $\low$ and a cardinality condition on $\sparse$, both $\low$  and $\sparse$ can be recovered exactly
with high probability by solving what is known as the program of principal pursuit (PCP):
 \begin{equation*}
\min_{\low,\sparse}\|\low\|_*+ \bar \gamma\|\sparse\|_1 \quad \text{
subject to } \quad
 \low+\sparse=\data
\end{equation*}
where $\bar \gamma=(\max(m,n))^{-1/2}$ is a regularization parameter. The output of the low rank component is referred to as robust principal components (RPC). Compared to the original problem,  the rank constraint on $\low $ and the cardinality constraint on $S$ have been replaced by convex functions.
The  incoherence condition
on   the singular vectors of $\low$  prevents
the  low rank component from being   too sparse. The cardinality condition requires the support of  $\sparse$ to be selected uniformly at random so that $\sparse $ is not low rank.\footnote
  {More precisely,  Theorem 1.1 of \citet{candes-li-ma-wright} shows that incoherent low-rank matrix can be recovered from non-vanishing fractions of gross errors in polynomial time.
The proof assumes  that $S$ is  uniformly
  distributed on the set of matrices with support of a fixed cardinality.}
 \citet{zlwcm}  allows for the presence of small noise  $\noise$. With  $Z=\low+\sparse+\noise$, it is shown that $\low$ and $\sparse$ can still be recovered
with high probability upon solving the convex program
\[ \min_{L,S} \norm{L}_*+\bar \gamma \norm{S}_1 \quad \text{s.t. } \quad \norm{Z-L-S}_F\le \delta
\]
  if, in addition to the incoherence and cardinality conditions,  $\|\noise\|\le
\delta$ holds for some known $\delta$. This result establishes that RPC is stable  in the presence of small entry-wise noise, a setup that seems appropriate for factor analysis. In summary, the main insight from matrix recovery problems is that while rank minimization is NP hard, the surrogate problem of nuclear-norm minimization still permits recovery of the desired low rank matrix with high probability.

In terms of implementation, singular value thresholding algorithm (SVT) plays an important role in the reparameterized problem. For a $m\times n$ matrix $Z$ and   $\textsc{svd}(Z) =U_rD_rV_r\pr+U_{n-r}D_{n-r} V_{n-r}^\pr $, define
\begin{equation}\soft_r^\gamma=\bigg[D_r-\gamma I_r\bigg]_+\equiv
\max(D_r-\gamma I_r,0).
\label{eq:soft}
\end{equation}
Importantly, the SVT is the proximal operator associated with the nuclear norm.\footnote{A proximal operator specifies the value that minimizes an approximation to a regularized version of a function. Proximal methods are popular in convex optimizations. See, for example, \citet{boyd-now}.}
 Theorem 2.1 of \citet{cai-candes-shen} shows that
\begin{eqnarray}
\label{eq:problem1}
 U_r D_r^\gamma V_r^\prime=\min_{\substack{L}}
\gamma\norm{L}_*+\frac{1}{2}\norm{\data-\low}_F^2.
\end{eqnarray}
 In other words, the optimal approximation of the low rank component of $Z$ under rank constraint is $U_rD_r^\gamma V_r^\prime$
where $D^\gamma_r$
is a matrix of thresholded singular values.  Compared with the unregulated estimate of $U_rD_rV_r^\prime$, the only difference is that the singular values are thresholded.
 It is possible for $D_r^\gamma$ to have rank $r^*<r$ because of thresholding. As a consequence, the rank of the regulated estimate of the low rank component can be smaller than the unregulated estimate.\footnote{Algorithms that solve (\ref{eq:problem1}) include Augmented Lagrange Multiplier and Accelerated Proximal Gradient (\citet{lin-chen-ma}), ADMM (\citet{boyd-now}), and CVX (\citet{cvx-guide}).}

\subsection{Relation to Factor Models}
The previous subsection provides results for recovery of the low rank and spares components, $\low$ and $\sparse$. Since $\low$ has rank $r$, it can be factorized as a product of two rank $r$ matrices, $A$ and $B$, that is, $L=AB^\prime$. This subsection discusses the recovery of $A$ and $B$. This is useful since we are eventually interested in the factors and the loadings of a minimum-rank factor model, not just the common component.

The key step that ties a low rank decomposition to factor analysis is to establish that
 the regularized problem with  $\gamma$ as threshold parameter, i.e.
\begin{equation}
\label{eq:problem2}
\min_{A,B} \frac{1}{2}\norm{Z-AB\pr}_F^2 +\gamma\norm{AB\pr}_*
\end{equation}
has solution
\begin{equation}
\label{eq:rpca-solution}
 \bar A=  U_r( \soft^\gamma_r)^{1/2}, \quad  \bar  B=  V_r (  \soft^\gamma_r)^{1/2},
\end{equation}
and that  $\bar L=\bar A\;\bar B^\prime$  also solves (\ref{eq:problem1}). The  result  that $(\bar A,\bar B)$ solves the problem (\ref{eq:problem2}) if and only if  $\bar\low=\bar A\;\bar B\pr$ solves (\ref{eq:problem1}) was first noted in \citet{rennie-srebro}. Detailed proofs are given in \citet{hmlz} and \citet{boyd-now}. A sketch of the idea is as follows.

Since $AB^\prime= U_r D_r V_r\pr$, and $U_r$ and $V_r$ are orthonormal, then by Cauchy-Schwarz inequality,
\begin{eqnarray*}
    \tr( D_r)&=&\tr( U_r\pr AB\pr  V_r  )
    \le  \norm{A}_F\norm{B}_F\\
    & \le & \frac{1}{2} \big(\norm{A}_F^2+\norm{B}_F^2\big).
\end{eqnarray*}
But since $L=AB^\prime$ by definition, it follows that $\norm{L}_*=\tr( D_r)$ and the above implies
\[\norm{L}_*\le \frac{1}{2} \big(\norm{A}_F^2+\norm{B}_F^2\big)\]
with equality when $ A=  U_rD_r^{1/2}$ and $   B=  V_r D_r^{1/2}$. Hence (\ref{eq:problem2})  is just a reformulation of (\ref{eq:problem1}) in terms of $A$ and $B$.  Consider now the first order conditions associated with (\ref{eq:problem2}).
If $\bar A$ and $\bar B$ are solutions, it must be that
$-(Z-\bar  A\bar  B')\bar  B+\gamma  \bar  A= 0$ and $ -(Z-\bar  A\bar  B')'\bar  A+\gamma  \bar  B = 0$.
Left multiplying the first condition by $\bar A\pr$ and the second by $\bar B \pr$, we see
 that $\bar A\pr \bar A=\bar B\pr \bar B$ when the first order conditions hold.
Rearranging terms, we obtain
\begin{eqnarray*}
	\begin{pmatrix} -\gamma  I & Z \\ Z' & -\gamma  I\end{pmatrix}
	\begin{pmatrix} \bar  A\\ \bar  B \end{pmatrix}&=&
	\begin{pmatrix}
		\bar  A \\ \bar  B \end{pmatrix}      \bar A\pr \bar A .
\end{eqnarray*}
This has the generic structure $\mathbb Z \mathbb V=\mathbb V \mathbb X$, which is an eigenvalue problem. In particular, the eigenvalues values of $\mathbb X$ are those of $\mathbb Z$, and $\mathbb V$ are the corresponding eigenvectors. In the present context,
the eigenvalues of $\bar A\pr \bar A$ are those of the first matrix on the left, and $(\bar A,\bar B)$ are the corresponding left and right eigenvectors. Though  $-\gamma \pm{ d_i}$ are two possible solutions for every $i$,  we only accept  $\sqrt{d_i-\gamma}$ for positivity.  Collecting the thresholded singular-values into $D_r^\gamma$,  the solution defined in (\ref{eq:rpca-solution}) obtains.
 These are the  robust principal components of $Z$ under the assumed normalization that $\bar  A'\bar  A=\bar  B'\bar  B=\bar D_r^\gamma$.\footnote{ \citet{uhzb} referred to (\ref{eq:problem2}) as {\em quadratically regularized PC}. See  their Appendix A for a complete proof.}

\section{Constrained Approximate Factor Models}

An important result in  large dimension factor modeling is that the factor space can be consistently estimated by principal components when $N$ and $T$ are large, as reviewed in Section 2. A key finding in the unsupervised learning literature  is that robust principal components can be obtained via SVT for a given sample of data, as reviewed in Section 3. One might expect RPC to play a role in rank constrained factor analysis. Indeed there is one, and this section makes this precise.  Given that RPC is a method developed in the machine learning literature, we start by clarifying  some differences between the statistical and the algorithmic approach to low rank  modeling.

Notably, the decomposition $\data=\low+\sparse$ is consistent with many probabilistic structures. Statistical factor analysis specifies one particular  representation:  $X=F\Lambda\pr + e$. We use Assumption A to restrict the factors, loadings, and the idiosyncratic noise  so that the eigenvalues of the common component  $C=F\Lambda\pr$ diverge with $N$. We  find $F$ and $\Lambda$ to minimize the sum of squared residuals and let $e$ be residually determined. We establish that $\Fpca$ is consistent at rate $\sqrt{N}$, $\Lpca$ at rate $\sqrt{T}$, and  $\hat C$ at rate  $\min(\sqrt{N},\sqrt{T})$ assuming that the factor model is correctly specified.

In contrast,  machine learning analysis is `distribution free' and the data generating process is left unspecified. Often,
  $\sparse$ is explicitly chosen together with $\low$, not residually determined.
For Netflix type problems when a low rank matrix  is to be recovered from data with missing values, the probability of exact recovery is typically obtained under  an incoherence condition that makes no reference to eigenvalues.   But  when the data are noisy rather missing,  one can only hope to recover an approximate (rather than the exact) low rank component.\footnote{For application of the incoherence condition in matrix completion, see \citet{candes-li-ma-wright}. For general matrix recovery problems, see \citet{agarwal-negahban-wainwright:12} and \citet{negahban-wainwright:12},  .} What matters most then is  the singular vectors associated with the large singular values.  For such problems, which seem more comparable to our setup, \citet{agarwal-negahban-wainwright:12}  obtains an error bound in the Frobenius norm  for  $\low$ and for $\sparse$ under the assumption that $\norm{L}_\infty\le \frac{\alpha}{\sqrt{mn}}$. But recall that $\norm{L}_\infty=\max_{i,t}|L_{it}|$ and    $ \norm{L}^2_F=\sum_{i=1}^m \sum_{t=1}^n |L_{it}|^2=\sum_{i=1}^r d_{L,i}^2$.  The condition $\norm{L}_\infty\le \frac{\alpha}{\sqrt{mn}}$ is effectively a restriction on the sum of the eigenvalues of the low rank component $L$.

What transpires from  the above discussion is that  machine learning methods restrict the sample eigenvalues  of $\low$ and obtain finite sample error bounds. In contrast,  approximate factor analysis puts restrictions on the population eigenvalues of the common component $C$ through moment conditions collected into Assumption A, which also enable precise parametric convergence rate to be obtained. Interestingly, Corollary 2 of \citet{agarwal-negahban-wainwright:12} suggests that for the spike mean model, the error of the low rank component is of order $\frac{N+T}{NT} \approx \min(N,T)^{-1}$ with high probability.  This agrees with the asymptotic convergence rate obtained in our previous work on the unconstrained case.

 A machine learning analysis closest in spirit to ours is \citet{bcm:16}. This paper reformulates estimation of minimum rank factor model for iid data as one of smooth optimization under convex compact constraints. Via lower bounds,  the solutions are shown to be certifiably optimal in many cases, requiring only that $\Sigma_X=\Sigma_C+\Sigma_e$, where  $\Sigma_e$ is diagonal. Our data need not be iid, and $\Sigma_e$ need not be diagonal, but we invoke more assumptions to provide parametric convergence rates. The results provide different perspectives to a related problem.

\subsection{Robust Principal Components (RPCA): ($\Frpca,\Lrpca)$}

This subsection considers estimation of $r^*$ factors,  or what was referred to in Section 3 as the approximate minimum rank. In some cases, economic theory may suggest $r$ factors but only $r^*<r$ may be empirically relevant. In other cases, economic theory may suggest $r^*$ factors, as in affine term structure models for interest rates. But financial data tend to have fat tails. Extreme values may lead us to the discovery of  more than $r^*$ factors from the data. As principal components are blind as to whether pervasive variations or extreme events underlie the components, it is desirable to have a way to guard against noise corruption.

One way to think about the noise-corruption problem is that the principal component estimates are not efficient when there is significant heterogeneity in the idiosyncratic errors.  \citet{boivin-ng-joe} considers re-weighting each series by the standard-deviation of the idiosyncratic error, obtained from a preliminary (unweighted) estimation of the factor model. A drawback is that these weights are themselves sensitive to outliers. A second approach is the POET estimator considered in \citet{fan-liao-mincheva}, which uses thresholding to enforce sparsity of the idiosyncratic errors of a spiked-covariance  matrix.

We propose to to apply thresholding to $\Sigma_C$ rather than $\Sigma_e$. This is appealing because the common component is the object of interest,  not the idiosyncratic errors. To do so requires that we map the scaled factor model given in (\ref{eq:scaled-model}) into the problem defined in (\ref{eq:problem2}). Consider the goal of recovering $C^*$ in the decomposition
\[Z=C+e=C^*+C^-+e \quad\quad C^*=F^*\Lambda^{*\pr}.\]
As in \citet{tenberge-kiers}, the common-component is decomposed into $C=C^*+C^-$. While $C$ has rank $r$,  it is well approximated by $C^*$ whose rank is $r^*$, and we want to estimate $C^*$ by the method of
robust principal components (RPC).\footnote{Motivated from an asset pricing perspective,   \citet{lettau-pelger:17} suggests to apply PCA to a covariance matrix of overweighted mean, or in our notation, $Z'Z+\gamma \bar Z\bar Z^\prime$, where $\bar Z$ is the mean of $Z$. Their risk-premium PCA also uses regularization to control for weak factors.} From the previous subsection, the rank regularized problem is
\begin{equation}
(\Frpca_z,\Lrpca_z)=\argmin_{F,\Lambda} \frac 1 2 \Big( \norm{Z-F\Lambda\pr}_F^2+\gamma\norm
{F}_F^2+\gamma \norm{\Lambda}_F^2\Big),
\label{eq:rapca-obj}
\end{equation}
with optimal solution,
\begin{equation} \label{eq:rapca-Z} \bar F_z= U_r (D_r^\gamma)^{1/2}, \quad \bar \Lambda_z =V_r (D_r^\gamma)^{1/2} \end{equation}
where $D^\gamma_r=[D_r-\gamma I_r]_+$.
As in the unconstrained case,  we work with normalized factors $\Frpca=\sqrt{T} \bar F_z$ and $\Lrpca=\sqrt{N} \bar \Lambda_z$:\footnote{These are also the optimal solutions from the following optimization problem
\begin{equation*}
(\Frpca,\Lrpca)=\argmin_{F,\Lambda} \Big(\frac{1}{NT}\norm{X-F\Lambda\pr}_F^2+ \frac{\gamma}{T}\norm
{F}_F^2+\frac{\gamma}{N}\norm{\Lambda}_F^2\Big).
%\label{eq:rapca-obj-X}
\end{equation*}
Though $(\bar F Q, \bar \Lambda Q)$ is also a solution for any orthonormal matrix $Q$,
$(\bar F,\bar \Lambda)$ as defined in (\ref{barF-X}) and (\ref{barL-X})  is the only solution (up to a column sign change) that satisfies $\frac{\Frpca\pr \Frpca}{T}=
\frac{\Lrpca\pr\Lrpca}{N} =\soft^\gamma _r$, assuming that the diagonal elements of $D_r^\gamma$ are distinct.
%Note that $D_r$ contains the first $r$ singular values of $X/\sqrt{NT}$.
 }
\begin{subequations}
\begin{eqnarray}
\label{barF-X} \bar F&=& \sqrt{T} U_r (D^\gamma_r)^{1/2}\\
\label{barL-X}
\bar \Lambda &=&\sqrt{N}V_r (D^\gamma_r)^{1/2}.
\end{eqnarray}
\end{subequations}
The restricted  and the unrestricted estimates are related by
\begin{eqnarray*}
   \Frpca &=& \Fpca \; \Delta_{NT}\\
   \Lrpca &=& \Lpca \; \Delta_{NT}
\end{eqnarray*}
where
\begin{equation}
\label{eq:Delta}
 \Delta_{NT}^2=D^\gamma_r D_r^{-1} =\diag\bigg(\frac {(d_1-\gamma)_+}{d_1},...,\frac{(d_r-\gamma)_+}{d_r}\bigg).
 \end{equation}
Hence while  $\frac{\Fpca\pr \Fpca}{T}=\frac{\Lpca\pr \Lpca}{N} =\soft_r$, now $\frac{\Frpca\pr \Frpca}{T}=
\frac{\Lrpca\pr\Lrpca}{N} =\soft^\gamma _r$.  Define  the rotation matrix for $\bar F$ by
   \[ \bar H_{NT} = \Hpca_{NT}  \Delta_{NT} \]
From the relationship between  $\Frpca$ and $\Fpca$,
\begin{equation}\label{bar-Ft} \bar F_t -\bar H_{NT}^{\prime} F_t^0 = \Delta_{NT}(\Fpca_t- \hat H_{NT}' F_t^0). \end{equation}
Hence $\bar F_t$ estimates a rotation of $F^0_t$.  But the inverse of $\bar H_{NT}$ is not the rotation matrix for $\bar \Lambda$.  As shown in  Appendix,
\begin{equation}\label{bar-Lambdai}
\bar \Lambda_i-\bar G_{NT} \Lambda_i^0 =
\Delta_{NT} [ \Lpca_i-\Hpca_{NT}^{-1} \Lambda_i^0]
\end{equation}
where the rotation matrix for $\bar \Lambda$ is
%\[ \bar G_{NT}=(D^\gamma_r)^{1/2} D_r^{-1} (D^\gamma_r)^{1/2} (\bar H_{NT}^\gamma)^{-1},\]
\[ \bar G_{NT}=\Delta_{NT}^2 (\bar H_{NT})^{-1}=\Delta_{NT}  \hat H_{NT}^{-1}\]

Denote the probability limit of $\Delta_{NT}$ as $\Delta_\infty= ( \mathbb D_r^\gamma \mathbb D_r^{-1})^{1/2}$,
where $\mathbb D_r^2$ is the diagonal matrix consisting of the eigenvalues of $\Sigma_\Lambda^{1/2} \Sigma_F\Sigma_\Lambda^{1/2}$, and $\mathbb D_r^\gamma =(\mathbb D_r-\gamma I_r)_+$.
Using Proposition \ref{prop:prop1}, (\ref{bar-Ft}), and (\ref{bar-Lambdai}), we obtain the following result.

\begin{proposition}
	\label{prop:prop2}
Let $(\Frpca,\Lrpca)$ be given in (\ref{barF-X}) and (\ref{barL-X}) with threshold parameter $\gamma>0$.
%Let $\mathbb D_r^\gamma=(\mathbb D_r-\gamma I_r)_+$.
Suppose that Assumption A holds and $N,T\rightarrow \infty$. Then
\begin{itemize}
    \item[(i)]
$	 \sqrt{N} (\Frpca_t- \bar H^{ \prime}_{NT} F_t^0) \dconv  N\bigg(0, \Delta_\infty\textsc{Avar}(\Fpca_t) \Delta_\infty\bigg)
	 \equiv N(0,\textsc{Avar}(\Frpca_t))$\\
\item[(ii)]  $ \sqrt{T} (\Lrpca_i -\bar G_{NT} \Lambda^0_i) \dconv  N\bigg(0, \Delta_\infty   \textsc{Avar}(\Lpca_i) \Delta_\infty \bigg) \equiv N(0,\textsc{Avar}(\bar \Lambda_i)).$
\end{itemize}
\end{proposition}
Since the diagonal elements of $\Delta_\infty$ are less than 1,
Proposition \ref{prop:prop2} implies that $\textsc{Avar}(\Frpca_t)\le \textsc{Avar}(\Fpca_t)$, and  $\textsc{Avar}(\Lrpca_i)\le
\textsc{Avar}(\Lpca_i)$.

Turning to the common component $C^0=F^0\Lambda^{0\prime}$, the RPC estimate is $\bar C =\bar F \, \bar \Lambda\pr$ and the PC estimates  is $\hat C=\Fpca \Lpca\pr$. Using $\Delta_{NT}^2$ defined in (\ref{eq:Delta}), we see that
\[ \bar C =\bar F \, \bar \Lambda'= \Fpca \Delta_{NT}^2 \Lpca\pr \ne \Fpca\Lpca\pr=\hat C. \]
Since $\hat C$ is an asymptotically unbiased estimate for the corresponding element of $C^0$, it follows that  the elements of $\bar C$ are biased towards zero. Recalling that $||\bar C||_F^2=\sum_{i=1}^N \sum_{t=1}^T \bar C_{it}^2=
\text{trace}(\bar C\pr \bar C)$,
% \[ \frac 1{NT} \tr(\bar C \bar C')=\tr((D_r^\gamma)^2) < \tr(D_r^2) = \frac 1 {NT} \tr(\hat C\pr \hat C).\]
\[ \frac { \tr(\bar C \bar C')}{\tr(XX')}
=\frac{\tr((D_r^\gamma)^2)}{\tr(D^2)} < \frac{\tr(D_r^2)}{\tr(D^2)} = \frac {\tr(\hat C\pr \hat C)}{\tr(XX')}.\]
Thus $\bar C$ accounts for a smaller fraction of the variation in $X$.

We can also derive the limiting distribution of the estimated common components.
Let $\bar C_{it}=\bar F_t'\bar \Lambda_i$ and $C_{it}^0=\Fp \Lambda_i^0$. As shown in the Appendix:
\begin{equation} \label{common-distribution}
 (A_{NT}(\bar C_{it}))^{-1/2}(\bar C_{it}-C_{it}^0-\text{bias} ) \dconv N(0,1) \end{equation}
where $\text{bias} = \gamma \Fp_t \bar H D_r^{-1} \bar H^{-1} \Lambda_i^0 $ and
\[ A_{NT}(\bar C_{it})= \frac 1 T \bar \Lambda_i' \textsc{Avar}(\bar F_t) \bar \Lambda_i + \frac 1 N \bar F_t' \textsc{Avar}(\bar \Lambda_i) \bar F_t. \]
The convergence rate is $\bar C_{it}$ is thus $\min\{\sqrt{N},\sqrt{T}\}$. Though we can remove the bias by setting $\gamma\rightarrow 0$,   $\gamma$ cannot be too small if we were to avoid over-estimating the dimension of the common component. Specifically, we need
$\gamma \gg \frac{1}{\min\{\sqrt{N},\sqrt{T} \}}$ so that the estimated number of factors would not be contaminated by the idiosyncratic errors. This is because the singular value of $\frac{e}{\sqrt{NT}}$ (or the square root of the largest eigenvalue of $\frac{e'e}{(NT})$ is no smaller than $O_p(\frac{1}{\min\{\sqrt{N},\sqrt{T}})$.

Though $\bar C$ is an asymptotically biased estimator for $C^0$, its asymptotic mean squared errors \textsc{Amse} may be smaller than that of the unbiased estimator $\hat C$.
To see why, consider $(i,t)$-th element:
$ \bar C_{it}= \bar F_t'\bar \Lambda_i  = \hat F_t' \Delta_{NT}^2 \hat \Lambda_i.$
Suppose that there is only a single factor $r=1$, then
\[\bar C_{it}=
\bigg(\frac{(d_1-\gamma)_+}{d_1}\bigg) \hat C_{it}\equiv \delta_1 \hat C_{it}.\]
The asymptotic bias and variance of $\bar C_{it}$ are, respectively,
\begin{eqnarray*}
 \textsc{Abias} (\bar C_{it}) &=& (\delta_1-1) C_{it}^0\\
\textsc{Avar}(\bar C_{it})&=& \delta_1^2 \textsc{Amse}(\hat C_{it}).
\end{eqnarray*}
 Thus the MSE of $\bar C_{it}$ is
$ \textsc{Amse}(\bar C_{it}) =(\delta_1-1)^2(C_{it}^0)^2 +\delta_1^2 \textsc{Amse}(\hat C_{it})$
so that
\[ \frac{\textsc{Amse}(\bar C_{it})}{{\textsc{Amse}(\hat C_{it})}}= (\delta_1-1)^2 \frac{(C_{it}^0)^2}{\textsc{Amse}(\hat C_{it})}+
\delta_1^2.
\]
As shown in \citet{bai-ecta-03},
the asymptotic MSE of $\hat C_{it}$ depends on $\Sigma_\Lambda$ and $\Sigma_F$, and on the variance of idiosyncratic errors.  But $\hat C_{it}$ is asymptotically unbiased for $C_{it}^0$, and  $\textsc{Amse}(\hat C_{it})=\textsc{Avar}(\hat C_{it})$.  Hence the relative \textsc{Amse} can be less than one when the signal of the common component is weak, which can be due to a small $\Sigma_\Lambda$ and/or a small $\Sigma_F$,  or when the idiosyncratic error variance is large. These correspond to cases of small singular values in the low rank component and noise corruption that motivated RPCA in machine learning. Here, we find that from a statistical perspective,  it is also beneficial to use the regularized principal components analysis when the data are noisy and when the pervasive signals are weak.

 It is noteworthy that soft-thresholding of the singular values is distinct from regularization of the singular vectors for a given rank of the low rank component, referred to in the literature as sparse principal components (SPC). The thresholding operation in SPC analysis does not change the rank of the factor estimates. It only performs shrinkage.\footnote{Sparse PCA is often motivated by easy interpretation of the factors. See, for example, \citet{shen-huang:08}.} In contrast, SVT constrains the rank of the low rank component to be no larger than $r$ as any factor corresponding to $d_i \le \gamma $ will effectively be dismissed. It has efficiency implications since  $\Dpca-\Dpca^\gamma\ge 0$ by construction. As a consequence, the asymptotic variance  $\Frpca_{jt}$ cannot exceed that of the unrestricted estimates $\Fpca_{jt}$ for all $j=1,\ldots r$.

\subsection{Selection of Factors}
The  problem defined in (\ref{eq:rapca-obj}) penalizes components with small contributions to the low rank component. Optimality is defined from an algorithmic perspective that does not take into account that $(\Frpca,\Lrpca)$ are estimates.  The large is $r$, the better is the fit, but variance also increases with the number of factors be estimated.  \citet{bai-ng-ecta-02} suggests to determine the  number of factors using criteria that take into account model complexity (hence sampling uncertainty).  These take the form
\[ \hat r=\min_{k=0,\ldots, \text{rmax}}\hat{ IC}(k), \quad\quad \hat {IC}(k)= \log(\textsc{ssr}_k) + k g(N,T).\]
The $\hat{IC_2}$ criterion, often used in empirical work, obtains when
\begin{equation}
g(N,T)=\frac{(N+T)}{NT}\log\bigg(\frac{NT}{N+T}\bigg). \label{eq:IC-02}
\end{equation}
The \textsc{ssr} term in the criterion function is the sum of squared residuals from fitting a model with $k$ factors. Suppose that each series in standardized, then $\norm{Z}_F^2=1$; together with $||\Fpca\Lpca\pr||=\norm{D_r}$, \textsc{ssr}$_k$ can be written as
\[\textsc{ssr}_k  =1-\sum_{j=1}^k d_j^2=\norm{Z-\Fpca_k\Lpca_k^\prime}_F^2.\]
In the constrained problem  (\ref{eq:rapca-obj}) yields $\norm{\Frpca\;\Lrpca\pr}_F^2=\norm{D_r^\gamma}_F^2$. For given $k$ and $\gamma>0$,
\[ \textsc{ssr}_k(\gamma)= 1-\sum_{j=1}^k (d_j-\gamma)_+^2=\norm{Z-\Frpca_k\Lrpca_k^\prime}_F^2.\]
This suggests to  define a class of criteria  that takes into account both the rank of the common component and sampling uncertainty as follows:
\begin{equation}
\bar r=\min_{k=0,\ldots,\text{rmax}}
\log\bigg(1-\sum_{j=1}^k (d_j-\gamma)_+^2\bigg) + k g(N,T) . \label{eq:IC-17}
\end{equation}
In other words, \textsc{ssr}$_k$ is evaluated at the rank restricted estimates $(\Frpca,\Lrpca)$.
Taking the approximation $\log(1+x)\approx x$, we see that
\[ \overline{ IC}(k)=\hat{IC}(k)+\gamma \sum_{j=1}^k\frac{(2d_j-\gamma)}{\hat{\textsc{ssr}}_k}.
\]
Since $d_j\ge d_j-\gamma\ge 0$, the penalty is heavier in $\overline{IC}(k)$ than $\hat {IC}(k)$. The rank constraint adds a data dependent term to each additional factor, hence  a more conservative estimate of $r$.
An appeal of the criterion is that the data-dependent adjustment does not require the researcher to be precise about the source of the small singular values. They can be due to genuine weak factors, noise corruption, omitted lagged and non-linear interaction of the factors that are of lesser importance.

The larger is $\gamma$, the stronger is the penalty. A natural question to ask is how to determine $\gamma$. Since $X$ is standardized  and $Z=\frac{X}{\sqrt{NT}}$,  we have $\norm{Z}_F^2=1$ by construction. Thus $ |d_j|^2$ is the total variation of $Z$ that the $j$-th unconstrained factor will explain. Regularization reduces its contribution in a  non-linear way.  In applications, we set $\gamma$ to 0.05. If $d_1=0.4$,  the contribution of $F_1$  will fall from $.4^2=0.16$ to $(.4-.05)^2=.1225$. If $d_2=0.2$, the contribution of $F_2$ will fall from 0.04 to 0.025, a larger percentage drop than $F_1$. Any  $d_j<.05$  will be truncated to zero. While  $\gamma=0.05$ is a small penalty to $d_j$, $\sum_{j=1}^k (2d_j-\gamma)$ makes a non-trivial difference to  the determination of $r$ as we will see in simulations as we will see below.

\subsection{Practical Considerations}

 Consider the infeasible linear regression  $y_{t+h}= \alpha^\prime F_t+\beta^\prime W_t+\epsilon_{t+h}$ where $W_t$ are observed predictors.
The regression is infeasible because $F$ is latent.    \citet{bai-ng-ecta-06} shows that  the APC estimate $\tilde F$ can be used in place of the latent $F$, and inference can proceed as though $F$ was known  provided $\frac{\sqrt{T}}{N}\rightarrow 0$. Section 2 shows that it is equally valid to replace $F$ with the PC estimate $\Fpca$.

What happens if we replace $F$ by the RPC estimate $\Frpca$? It may  be tempting to think that $\Frpca$ will reduce the goodness of fit because $\var(\Frpca)<\var(\Fpca)$. This, however, is not true.
Least squares estimation will give identical fit whether we use $\Fapca, \Fpca, $ or $\Frpca$ as regressors. To appreciate this point, there are three cases to consider. First, suppose that SVT shrinks but does not threshold the singular values so that
$\Frpca $ and $\Fpca$ have the same rank.
Then $\Fapca$,  $\Fpca$ and $\Frpca$ are all spanned by $U_r$. In other words, they are perfectly correlated.  The   estimates of $\alpha$ will simply adjust to compensate for the difference in scale.    In the second case when $r^*=\text{dim}(\Frpca)<\text{dim}(\Fpca)=r$ because of thresholding, the fit provided by $\Frpca$ remains identical to the fit provided by the first $r^*$ columns of $\Fpca$ by virtual of the fact the omitted factors are orthogonal to the included ones.  However, when the predictive regression is augmented with $k\ge r$ factors,   a ridge regression of $y_{t+h}$ on $\bar F_t$ will not give the same estimates as a ridge regression of $y_{t+h}$ on the $k$ unregulated factors  $\hat F_{t}$ because no thresholding is involved in $\frac{\hat F^\prime \hat F}{T}=D_k$. With thresholding, $D_k^\gamma$ can expected to be reduced rank.

The upshot of the above discussion is that use of the rank restricted estimate $\Frpca$ will make no difference to factor augmented regressions unless an effort is made to shrink the coefficients associated with the factors towards zero. This can be achieved by replacing  least squares with ridge regressions. As an example, suppose that $W_t$ is empty and we regress $y_t$ on a vector of $r$ regulated factors $\bar F_t$ only. Let $\bar \alpha_{OLS}$ and $\bar \alpha_R$ be the OLS and ridge estimator for $\alpha$ respectively. For given $\kappa$ and $\kappa_T=\frac{\kappa}{T}$,
\begin{eqnarray*}
\bar\alpha_{ols}&=& (\Frpca\pr\Frpca)^{-1}\Frpca y
=(D^\gamma_r)^{-1/2} U' y/\sqrt{T}\\
\bar \alpha_{R} &=& (\Frpca\pr\Frpca+\kappa I_r)^{-1}\Frpca y \\
&=&(T D^\gamma_r+\kappa I_r)^{-1}(D^\gamma_r)^{1/2}\sqrt{T} U' y
= (D^\gamma_r+\kappa_T I_r)^{-1} (D^\gamma_r)^{1/2} U_r y/\sqrt{T}\\
&=& (D^\gamma_r+\kappa_T I_r)^{-1} D^\gamma_r \bar\alpha_{OLS} =(I_r+\kappa_T (D^\gamma_r))^{-1} \, \bar\alpha_{OLS}\\
& \approx & (I_r-\kappa_T D_r^\gamma)\bar\alpha_{OLS}.
\end{eqnarray*}
While  SVT implements the rank constraint on $\bar F$ via $D_r^\gamma$, a ridge regression shrinks $\hat\alpha_{OLS}$ towards zero so that the contribution of $ \bar \alpha_R^\prime \Frpca_t $ is smaller than $\bar\alpha_{OLS}^\prime \Frpca_t$.
 Note that $\bar\alpha_R$ can be constructed from the (i) least squares estimator $\bar \alpha_{OLS}$; (ii) the matrix of singular values $D$  of $Z$;  and (iii) the regularization parameters $\gamma$ and $\kappa$. Explicit ridge estimation is actually not necessary.   The optimal number of factors in predictive regression can be determined by the AIC or BIC.  The penalty on model complexity will be heavier than in the OLS case since the sum-of-squared residuals will depend on $\gamma$ and $\kappa$ in a data dependent way, analogous to the $\overline{IC}$ criteria.

The second practical issue concerns construction of the factors. Whether we are interested in the APC, PC, or RPC estimates, the singular vectors $U_r$ required. Many numerical methods are available to compute singular vectors when $Z$ is large in dimension. To compute  the first singular nvector, the method of {\em  power iteration} starts from an initial vector that has a non-zero component in the direction of the target singular vector. It then  recursively updates and re-normalizes the vector till convergence. The resulting vector is the largest singular vector. The idea can be extended to  compute the invariant subspace of all  singular vectors. As suggested in \citet{hmlz},  the algorithm can be modified to construct robust principal components. The following algorithm is from \citet{hmlz}.\footnote{Our algorithm differs only in that we do
\textsc{svd} of $F$ and $\Lambda$ in Steps (ii) and (iv) instead of $FD$ and $\Lambda D$.}

\paragraph{Algorithm RPC (Iterative Ridge):} Given a $m\times n$ matrix $Z$,   initialize a $m\times r$ matrix $F=\mathbb U\mathbb D$  where $\mathbb U$ is orthonormal and     $\mathbb D=I_r$.
\begin{itemize}
\item[A.]  Repeat till convergence
\begin{itemize}
    \item[i.]  (solve $\Lambda$ given $F$):  $\tilde{\Lambda}=Z\pr F (F\pr F+\gamma I_r)^{-1}$.
    \item[ii] (orthogonalize): Do \textsc{svd}$(\tilde {\Lambda })=\tilde{\mathbb U}_\Lambda\tilde{\mathbb D}_\Lambda \tilde{\mathbb V}_\Lambda\pr$ and
    let $\Lambda=\tilde{\mathbb U}_\Lambda\tilde{\mathbb D}_\Lambda$ and $\mathbb D=\tilde{\mathbb D}_\Lambda$.
    \item[iii] (solve $F$ given $\Lambda$): $\tilde{F}=Z \Lambda(\Lambda\pr \Lambda+\gamma I_r)^{-1}$.
    \item[iv] (orthogonalize): Do $\textsc{svd}(\tilde F)=\tilde{\mathbb U}_F\tilde{\mathbb D}_F\tilde{\mathbb V}_F\pr$ and
    let $F=\tilde{\mathbb U}_F\tilde{\mathbb D}_F$ and $ \mathbb D=\tilde{\mathbb D}_F$.
\end{itemize}
\item[B.] (Cleanup) From
 $\textsc{svd}(Z\mathbb U_\Lambda)=U_rD_r\mathbb V\pr$, let $V_r=\mathbb U_\Lambda \mathbb V$, $D_r^\gamma=(D_r-\gamma I_r)_+$.
 \end{itemize}
 Algorithm RPC uses iterative ridge regressions to construct the factors and the loadings. The two \textsc{svd} steps ensure that the factors and loadings are mutually orthogonal.
The converged result of Step A gives $(\bar F_z, \bar \Lambda_z)$,  which is the solutions to the nuclear norm minimization problem stated in (\ref{eq:rapca-obj}). In theory, this is all that is needed for construction of $\Frpca=\sqrt{T} F_z$ and $\Lrpca=\sqrt{N} \Lambda_z$. But improved estimates of the left and right singular vectors can be obtained using Step B, which also explicitly thresholds the singular values.\footnote{In principle, the converged $\tilde{\mathbb U}_\Lambda$ should be $V_r$. Step B essentially computes an improved estimate by performing a \textsc{svd} of $Z\mathbb U_\Lambda=U_rD_r V_r\pr \mathbb U_\Lambda$, and then recovers $V_r$ from the right eigenvector of $Z\mathbb U_\Lambda$.
 } The final estimates of the factors and loadings that emerge from from step B are $\bar F_z= U_r (D_r^\lambda)^{1/2}$ and $\bar \Lambda_z =V_r (D_r^\lambda)^{1/2}$.
%from the first $r$ singular values of $Z$ (i.e., $D_r)$ and their right and left singular vectors $U_r$ and $V_r$, as well as $D_r^\gamma$.
The entire procedure only involves SVD for matrices of dimension $m\times r$ and $n\times r$, not dimension of $m\times n$. This is important when both $m$ and $n$ are large.
 When $Z$ is not huge in dimension, $(\bar F_z, \bar \Lambda_z)$ can be directly computed from an \textsc{svd} of $Z$,  and the algorithm is not necessary. The ridge regression perspective is nonetheless useful in highlighting the role that regularization plays in RPC.

\section{Linear Constraints}
The minimization problem in (\ref{eq:rapca-obj}) has a unique solution under the normalization
 $F\pr F =\Lambda\pr \Lambda =D_r $. However, the unique solution may or may not have economic interpretations.
This section considers
$m$ linear restrictions on $\Lambda$ of the form
\begin{equation} \label{eq:restriction} R \, \vec(\Lambda) = \phi   \end{equation}
 where $R$ is $m \times Nr$, and $\phi$ is $m\times 1$. Both $R$ and $\phi$ are assumed known  a priori. Economic theory may imply lower triangularity of the top $r\times r$ sub-matrix of $\Lambda$ when the data are appropriately ordered. By suitable design of $R$, the causality restriction can be expressed as $R\, \text{vec}(\Lambda)=\phi$ without ordering the data a priori.  Cross-equation restrictions are allowed, such as due to  homogeneity of the loadings across individuals or a subgroup of individuals suggested by theory. Other restrictions are considered in \citet{stock-watson:handbook-16}.
 The Appendix provides an example how to implement the restrictions in \textsc{matlab}.

The linear restrictions on the loadings considered here  contrasts  with sparse principal components  (SPC) estimation which either imposes  \textsc{lasso} type penalty on the loadings, or shrinks the individual entries to zero in a data dependent way, as  our constraints are known a priori.\footnote{For SPC, see  \citet{jolliffee-trendafilov-uddin}, \citet{ma:13}, \citet{shen-huang:08}, and \citet{zht}. The SPC is in turn different from the POET estimator of \citet{fan-liao-mincheva}  which constructs the principal components from a matrix that shrinks the small singular values towards zero.}
 Note that if we regard the diagonality of $F\pr F$ and $\Lambda\pr\Lambda$ as identification restrictions (rather than statistical normalizations), the linear  constraints on the loadings (\ref{eq:restriction}) constitute over-identifying restrictions with which we can use to test economic hypothesis. However, it is now possible to relax some of the diagonality restrictions, so long as they are replaced by a sufficient number of linear restrictions; identification is then still possible. A theory of identification for high dimensional factor models is given in   \citet{bai-wang-14}.

The constrained  factor estimates $(\Frpca_{\gamma,\tau},\Lrpca_{\gamma,\tau})$ are defined as solutions to the penalized problem
\begin{equation}
\label{eq:constrained-obj}
(\Frpca_{\gamma,\tau},\Lrpca_{\gamma,\tau})=\min_{F,\Lambda} \frac{1}{2} \norm
{Z-F\Lambda'}_F^2+ \frac{\gamma }{2}\bigg(\norm{F}_F^2+\norm
{\Lambda}_F^2\bigg)+\frac{\tau}{2} \norm{R \, \vec(\Lambda) -\phi}_2^2
\end{equation}
where $\gamma$ and $\tau$ are regularization parameters.    The linear constraints can be imposed with or without the rank constraints. Imposing cross-equation restrictions will generally require iteration till the constraints are satisfied.

 The first order condition with respect to $F$ for a given $\Lambda$ is unaffected by the introduction of the linear constraints on $\Lambda$. Hence, the solution
\begin{equation}
\label{eq:Fc-given-L}
\bar F_{\gamma,\tau} = Z \Lambda(\Lambda\pr \Lambda+\gamma I_r)^{-1},\quad \forall\tau \ge 0
\end{equation}
 can be obtained from a ridge regression of $Z$ of $\Lambda$.
\
To derive the first order condition with respect to $\Lambda$, we rewrite the problem in vectorized form:
\[ \|Z-F\Lambda'\|^2_F=\|\vec(Z')-(F\otimes I_N)\vec(\Lambda)\|_2^2, \quad \|\Lambda\|_F^2 =\|\vec(\Lambda)\|_2^2. \]
The first order condition with respect to $\vec(\Lambda)$ is
\begin{eqnarray*}
0&=& -(F'\otimes I_N)\Big[\vec(Z')-(F\otimes I_N)\vec(\Lambda)\Big]+ \gamma \, \vec(\Lambda) + \tau R'[R \, \vec(\Lambda)-\phi] \\
&=& -\vec(Z'F)-\tau R'\phi +(F'F\otimes I_N) \, \vec(\Lambda) + \gamma  \,\vec(\Lambda ) +\tau R'R \,\vec(\Lambda).
\end{eqnarray*}
Solving for $\vec(\Lambda)$ and  and denoting the solution by $\vec(\bar \Lambda_{\gamma,\tau})$, we  obtain
\begin{eqnarray}  \label{additional_penalty}
\Lrpca_{\gamma ,\tau} &=&\Big((F'F\otimes I_N) +\gamma  I_{Nr} +\tau  R'R \Big)^{-1} \Big[ \vec(Z'F)+\tau R'\phi\Big] \\
&=&\Big((F'F +\gamma I_r)\otimes I_N +\tau R^\prime R\Big)^{-1}\bigg[ \vec(Z'F+\tau R^\prime \phi)\bigg] \nonumber
\end{eqnarray}
where the last line follows from the fact that $(F'F\otimes I_N) +\gamma  I_{Nr}=(F'F+\gamma  I_r) \otimes I_N$.
Equations (\ref{eq:Fc-given-L}) and (\ref{additional_penalty}) completely characterize the solution under rank and linear restrictions. In general, the solution will need to be solved by iterating the two equations until convergence. A reasonable starting value is $(\Frpca,\Lrpca)$, the solution satisfying the rank constraint and before the linear restrictions are imposed. However, while $\Frpca\pr \Frpca=\Lrpca\pr \Lrpca=D_r^\gamma$ and $D_r^\gamma$ is diagonal, $\bar F_{\gamma,\tau}^\prime \bar F_{\gamma,\tau}$ and $\bar \Lambda_{\gamma,\tau}^\prime \bar \Lambda_{\gamma,\tau} $ will  not, in general, be diagonal when linear restrictions are present.
% matricesmutual orthogonality of the factors and the loadings should not be expected to hold when the linear restrictions are imposed.

These constraint will not bind unless $\tau=\infty$, and  we denote by $\Lambda_{\gamma,\infty}$ the binding solution. Observe that  in the absence of linear constraints (i.e. $\tau=0$),
\begin{equation} \label{eq:ridge}
  \vec(\Lridge) =\Big((F'F +\gamma I_r)\otimes I_N \Big)^{-1} \vec(Z'F) \end{equation}
which is a ridge  estimator. Furthermore, (\ref{eq:Fc-given-L}) and (\ref{eq:ridge}) are nothing but the RPCA estimates. The following is an estimator  that satisfies  both the rank constraint and $R \, \vec(\Lambda)=\phi$.
\begin{proposition}
\label{prop:prop3}
For given $F$, let $\Lcpca$ be the solution to (\ref{eq:constrained-obj}) with $\tau=\infty$. Also let $\Lridge$ be the solution  with $\tau=0$.
Then two solutions are related as follows:
\begin{equation} \label{RR-simplified}
\vec(\Lcpca)=\vec(\Lridge)-
[(F'F+\gamma  I_r)^{-1}\otimes I_N] R'
 \cdot  \Big[ R[(F'F+\gamma  I_r)^{-1}\otimes I_N] R'\Big]^{-1}\Big( R \, \vec(\Lridge)-\phi\Big)\end{equation}
\end{proposition}
Proposition \ref{prop:prop3} says that for given $F$, a restricted estimate of $\Lambda$ that satisfies both the rank and linear restrictions can be obtained by imposing the linear restrictions on RPCA solution of $\Lambda$ that only imposes rank restrictions.
It is easy to verify $\Lcpca$ satisfies  restriction (\ref{eq:restriction}). Once the restricted estimates are obtained, $F$ needs to be re-estimated based on (\ref{eq:Fc-given-L}). The final solution is obtained by iterating
(\ref{eq:Fc-given-L}) and (\ref{RR-simplified}).  We note again that $\bar F_{\gamma,\infty}^\prime \bar F_{\gamma,\infty}$ and $\bar \Lambda_{\gamma,\infty}^\prime \bar \Lambda_{\gamma,\infty} $ will not, in general, be diagonal matrices in the presence of linear restrictions.

Finally, a more  general  regularized problem we can consider is:
\[ (\Frpca_{\gamma_1,\gamma_2,\tau},\Lrpca_{\gamma_1,\gamma_2,\tau})=
\argmin_{F,\Lambda} \bigg(\frac{1}{2} ||Z-F\Lambda^\prime||_F^2
+\frac{\gamma_1}{2} ||F||_F^2 +\frac{\gamma_2}{2}||\Lambda||_F^2+\frac{\tau}{2
}||R\vec(\Lambda)-\phi)||_2^2\bigg). \]  Let \[ \bar D_r^{\gamma}=
(D_r-\sqrt{\gamma_1\gamma_2}\, I_r)_+.\] Relaxing the constraint that
$\gamma_1=\gamma_2$ with $\tau=0$ yields the general solution
\begin{subequations}
\begin{eqnarray} \bar
F_{\gamma_1,\gamma_2,\tau=0}&=&\Big(\frac{\gamma_2}{\gamma_1}\Big)^{1/4} U_r
(\bar D_r^\gamma) ^{1/2} \label{generalize-RPCA-F}\\ \bar
\Lambda_{\gamma_1,\gamma_2,\tau=0}&=&
\Big(\frac{\gamma_1}{\gamma_2}\Big)^{1/4}  (\bar D_r^\gamma)^{1/2}. \label
{generalize-RPCA-L} \end{eqnarray}
\end{subequations}
The corresponding common
component is \[ \bar C_{\gamma_1,\gamma_2,0}= U_r \bar D_r^\gamma V_r^\prime\]
When $\gamma_1=\gamma_2=\gamma$, the solution coincides with (\ref{eq:rapca-Z}), which can be computed by Algorithm RPC. Once the rank constrained solutions are obtained, Proposition
\ref{prop:prop3} can be used to impose linear constraints ($\tau=\infty$). For
other values of $\tau$, optimal solutions can be obtained via iterated ridge
regressions, for which $\gamma$ is replaced by $\gamma_1$  in equation
(\ref{eq:Fc-given-L}), and by $\gamma_2$ in equation
(\ref{additional_penalty}).

\section{Simulations and Application}

A small  simulation exercise is used to highlight the issues. Data are generated according to
\[ X_{it}=F^{0\prime}_t\Lambda_i^{0}  + e_{it} + s_{it}, \quad \quad e_{it}\sim (0,1) \]
where the sparse error $s_{it}\sim N(\mu,\omega^2)$ if $(i,t)\in \Omega$ and zero otherwise, $\Omega$ is an index set containing $(i,t)$ positions with non-zero values of $s_{it}$.  It is assumed a fraction $\kappa_N $  of cross-section units have outliers in a fraction $\kappa_T $  of the sample. In the simulatios, we let $(\kappa_N,\kappa_T)=(0.1,0.03)$, $\mu=5$.  Two data generating processes both with $r=5$ are considered.
\begin{itemize}
  \item DGP1: $F_t^0\sim N(0,I_r)$, $\Lambda_i^0\sim N(0,I_r)$, with $\omega\in (5,10,20)$;
  \item DGP2: $F^0=U_rD_r^{1/2}$, $\Lambda^0=V_rD_r^{1/2}$, with $\diag(D_r)=[1,0.8,0.5,0.3,0.2\theta]$, and $\omega=5$. Three values of $\theta$ are considered:  $(1,0.75,0.5)$.
\end{itemize}
The first DGP is designed to study the effect of outliers, which is expected to lead to an over-estimation of $r$. The second DGP varies the contribution of the smallest factor by the parameter $\theta$. The minimum rank is expected to decrease with $\theta$. We define $r^*$ to be the number of factors that contributes at least a fraction $c(s)=\frac{\var(S)}{\var(X)}$ of the variance of the common component of $X$. Note that $r^*$ is also the rank of $D_r^\gamma$. With $c(S)=0.05$, DGP has $r^*=5$ and DGP 2 has $r^*=3$.

The properties of the factor selection rules depend on the strength of the factors as well as the extent of noise contamination. We summarize these features using three statistics. The first is $C^r$, which  denotes the fraction of population variance $X$ due to all $r$ factors. The second is  $C_r$, which denotes the fraction of variance due to $r$-th (i.e. smallest) factor. These two indicate the relative importance of the common component  and the smallest factor in the data, respectively.   The third is $c(S)$, which denotes the fraction of variance of $Z$ due to the outliers in $S$. We report the mean of $\hat r$ and $\bar r$, the probability that $\hat r=r$ and $\bar r=r$ in 5000 replications with $rmax$ set to 8. To evaluate how well the estimated factors approximate the space spanned by the true factors,  we regress the smallest factor estimated on all $r^*$ singular vectors of the true common component. If the factor space is precisely estimated, the $R^2$ of this regression should be close to one. These are denoted $\hat R^2$ and $\bar R^2$ for the PC and RPC estimates of $F$, respectively.

Table 1 shows that in the absence of outliers, i.e. $c(S)=0$, the IC performs well and always correctly selects $r=5$ factors when $N$ and $T$ are both reasonably large. Rank regularization does not affect the number of factors selected in this setting. However, when noise corruption is present and $c(S)>0$ , $\hat r$ tends to exceed $r$ and has a mean of over 6. The higher is $c(S)$, the larger is the contribution of the outliers, and the larger is $\hat r$. However, $\bar r$ is more robust and correctly selects five factors in many cases.

Table 2 shows results for which has $r=5$ but $r^*=3$. This means that the smallest two  factors contribute less than $c(S)=0.05 $ of the variation in $C$. Even in the absence of measurement noise, $\hat r$ has a mean of four, implying that it tends to accept at least one of the small factors as valid. In contrast, $\bar r$ which has a mean of three tends to disregard both small factors. When measurement errors are allowed as in the bottom panel, $\hat r$ tends to estimate find an additional factor compared to the top panel. In contrast, $\bar r$ is unaffected by noise contamination. Of course, for this DGP, the true factor is $r$ and one can argue that $\hat r$ yields the correct estimate. As there is a tension between consistent estimation of $r$ and parsimony, it is up to the user whether to use $\hat r$ or $\bar r$. In applications, researchers tend to focus on the dominant ones. Our $\overline{IC}_2$ criterion provides a way to determine the minimum number of factors that should be used.

%% test3.m
%%% test3bal.m
We also estimate the factors using data from FRED-MD (\citet{fred-md}),  a macroeconomic  database consisting of a panel of 134 series over the sample 1960M1-2016M08. Consistent with previous studies, some series are transformed by taking logs and first differencing before the  factors are estimated. The panel is not balanced, we use the EM algorithm suggested  in \citet{stock-watson-di} which imputes the missing values from the factor model and iterate till convergence. The unbalance panel has $N=128$ variables with $T=676$ observations. We also consider a  balanced panel with $N=$ 92 series.\footnote{One idea is to use matrix completion algorithms to fill the missing values. But the data  are not `missing at random', and we leave it for future work.}

The squared-singular values  can be interpreted as the percent contribution of the factor to the variation of $Z$. As noted above, regularization shrinks the singular values and hence the length of the factors and the loadings towards zero.  Hence the unregularized singular values $\bar d_i^2$ are always smaller than the unregularized ones by roughly $\gamma^2=(0.05)^2$. The original $\hat {IC}_2$ finds eight factors for the balanced  panel. After regularization, $\overline {IC}_2$ finds three factors. In \citet{gorodnichenko-ng:17}, this difference between $\hat r$ and $\bar r$ is attributed to  interactions of the level factors disguising as separate factors. Instability in the loadings, along with outliers may also contribute to the difference.
 We then use  eight factors to impute missing values in the non-balanced panel. The $\overline{ IC}_2$ criterion continues to find three factors in the resulting balanced panel. In this data, the first factor loads heavily on real activity variables, the second on interest rate spreads, and the third on prices.

 \begin{center}
Eigenvalues of FredMD data
\vspace{0.1in}

\begin{tabular}{l|ll|llll}
F & \multicolumn{2}{c}{Balanced Panel}  &  \multicolumn{2}{c}{Non-Balanced Panel}\\ \hline\hline
 & $\hat d_1^2 $ & $\bar d_1^2$ & $\hat d_1^2 $ & $\bar d_1^2$ \\ \hline \hline
1&0.1828 &  0.1426 & 0.1493 & 0.1131\\
2&0.0921 &  0.0643 & 0.0709 & 0.0468\\
3&0.0716 &  0.0473 & 0.0682 & 0.0446\\
4&0.0604 &  0.0384 & 0.0561 & 0.0349\\
5&0.0453 &  0.0265 & 0.0426 & 0.0245\\
6&0.0416 &  0.0237 & 0.0341 & 0.0182\\
7&0.0301 &  0.0152 & 0.0317 & 0.0164\\
8&0.0287  &  0.0143 & 0.0268 & 0.0129\\ \hline
$r^*$ & 8 &  3 & 8 &  3 \\ \hline\hline
\end{tabular}
\end{center}

\section{Conclusion}
This paper considers estimation of approximate factor models by regularized principal components with focus on two problems.  The first problem is rank regularization with RPCA as output. This is useful when the idiosyncratic errors have large singular values such as due to extreme outliers, or when some factors have small singular values, such as when the loadings are small. A new class of factor selection criteria is proposed that will give more conservative estimates when the strong factor assumption is questionable.  The second problem is linear restrictions such as motivated by economic theory. We show that the  solution is a transformation of the unrestricted estimates.  Our analysis provides a statistical view of matrix recovery algorithms and complements results in the machine learning literature.
\newpage
% mc1.m produces mc1.out
\begin{center}
Table 1: DGP 1

\vspace{0.1in}
\begin{tabular}{llll|ll|l|ll|ll|ll|ll}
 \multicolumn{4}{c}{parameters} & \multicolumn{2}{c}{signal} & noise & \multicolumn{2}{c}{mean} &
 \multicolumn{2}{c}{prob. $=r$} & \multicolumn{2}{c}{prob. $=r^*$} & \multicolumn{2}{c}{spanning}\\
 \hline
 $N$ & $T$ & $r^*$ & $\omega$ & $C^r$  & $C_r$ & $c(S)$ & $\hat r$ & $\bar r$ & $\hat r$ &
 $\bar r$ & $\hat r$ & $\bar r$ & $\hat R^2$ & $\bar R^2$\\ \hline
  100 & 100 & 5 & 5.00 & 0.83 & 0.12 & 0.00 & 5.00 & 5.00 & 1.00 & 1.00 & 1.00 & 1.00 & 0.98 &  0.98  \\
 100 & 100 & 5 &10.00 & 0.83 & 0.12 & 0.00 & 5.00 & 5.00 & 1.00 & 1.00 & 1.00 & 1.00 & 0.98 &  0.98  \\
 100 & 100 & 5 &20.00 & 0.83 & 0.12 & 0.00 & 5.00 & 5.00 & 1.00 & 1.00 & 1.00 & 1.00 & 0.98 &  0.98  \\
 100 & 200 & 5 & 5.00 & 0.83 & 0.13 & 0.00 & 5.00 & 5.00 & 1.00 & 1.00 & 1.00 & 1.00 & 0.98 &  0.98  \\
 100 & 200 & 5 &10.00 & 0.83 & 0.13 & 0.00 & 5.00 & 5.00 & 1.00 & 1.00 & 1.00 & 1.00 & 0.98 &  0.98  \\
 100 & 200 & 5 &20.00 & 0.83 & 0.13 & 0.00 & 5.00 & 5.00 & 1.00 & 1.00 & 1.00 & 1.00 & 0.98 &  0.98  \\
 100 & 400 & 5 & 5.00 & 0.83 & 0.13 & 0.00 & 5.00 & 5.00 & 1.00 & 1.00 & 1.00 & 1.00 & 0.98 &  0.98  \\
 100 & 400 & 5 &10.00 & 0.83 & 0.13 & 0.00 & 5.00 & 5.00 & 1.00 & 1.00 & 1.00 & 1.00 & 0.98 &  0.98  \\
 100 & 400 & 5 &20.00 & 0.83 & 0.13 & 0.00 & 5.00 & 5.00 & 1.00 & 1.00 & 1.00 & 1.00 & 0.98 &  0.98  \\
  50 & 100 & 5 & 5.00 & 0.83 & 0.10 & 0.00 & 5.00 & 4.95 & 1.00 & 0.95 & 0.00 & 0.05 & 0.95 &  0.95  \\
  50 & 100 & 5 &10.00 & 0.83 & 0.10 & 0.00 & 5.00 & 4.95 & 1.00 & 0.95 & 0.00 & 0.05 & 0.95 &  0.95  \\
  50 & 100 & 5 &20.00 & 0.83 & 0.10 & 0.00 & 5.00 & 4.95 & 1.00 & 0.95 & 0.00 & 0.05 & 0.95 &  0.95  \\
  50 & 200 & 5 & 5.00 & 0.83 & 0.11 & 0.00 & 5.02 & 5.00 & 0.98 & 1.00 & 0.00 & 0.00 & 0.93 &  0.96  \\
  50 & 200 & 5 &10.00 & 0.83 & 0.11 & 0.00 & 5.02 & 5.00 & 0.98 & 1.00 & 0.00 & 0.00 & 0.93 &  0.96  \\
  50 & 200 & 5 &20.00 & 0.83 & 0.11 & 0.00 & 5.02 & 5.00 & 0.98 & 1.00 & 0.00 & 0.00 & 0.93 &  0.96  \\
  50 & 400 & 5 & 5.00 & 0.83 & 0.11 & 0.00 & 5.05 & 5.00 & 0.95 & 1.00 & 0.95 & 1.00 & 0.91 &  0.96  \\
  50 & 400 & 5 &10.00 & 0.83 & 0.11 & 0.00 & 5.05 & 5.00 & 0.95 & 1.00 & 0.95 & 1.00 & 0.91 &  0.96  \\
  50 & 400 & 5 &20.00 & 0.83 & 0.11 & 0.00 & 5.05 & 5.00 & 0.95 & 1.00 & 0.95 & 1.00 & 0.91 &  0.96  \\
 \hline\hline
 100 & 100 & 5 & 5.00 & 0.81 & 0.12 & 0.02 & 5.36 & 5.00 & 0.64 & 1.00 & 0.64 & 1.00 & 0.63 &  0.98  \\
 100 & 100 & 5 &10.00 & 0.78 & 0.12 & 0.06 & 5.79 & 5.00 & 0.28 & 1.00 & 0.28 & 1.00 & 0.28 &  0.98  \\
 100 & 100 & 5 &20.00 & 0.69 & 0.12 & 0.17 & 6.81 & 5.00 & 0.00 & 1.00 & 0.00 & 1.00 & 0.01 &  0.97  \\
 100 & 200 & 5 & 5.00 & 0.81 & 0.13 & 0.02 & 5.67 & 5.00 & 0.33 & 1.00 & 0.33 & 1.00 & 0.32 &  0.98  \\
 100 & 200 & 5 &10.00 & 0.78 & 0.13 & 0.06 & 5.91 & 5.00 & 0.19 & 1.00 & 0.19 & 1.00 & 0.19 &  0.98  \\
 100 & 200 & 5 &20.00 & 0.69 & 0.13 & 0.17 & 7.13 & 5.00 & 0.00 & 1.00 & 0.00 & 1.00 & 0.00 &  0.98  \\
 100 & 400 & 5 & 5.00 & 0.81 & 0.13 & 0.02 & 5.88 & 5.00 & 0.12 & 1.00 & 0.12 & 1.00 & 0.12 &  0.98  \\
 100 & 400 & 5 &10.00 & 0.78 & 0.13 & 0.06 & 5.90 & 5.00 & 0.17 & 1.00 & 0.17 & 1.00 & 0.16 &  0.98  \\
 100 & 400 & 5 &20.00 & 0.69 & 0.13 & 0.18 & 7.15 & 5.00 & 0.00 & 1.00 & 0.00 & 1.00 & 0.00 &  0.98  \\
  50 & 100 & 5 & 5.00 & 0.81 & 0.10 & 0.02 & 5.32 & 4.92 & 0.68 & 0.92 & 0.00 & 0.08 & 0.65 &  0.95  \\
  50 & 100 & 5 &10.00 & 0.78 & 0.10 & 0.06 & 5.69 & 4.89 & 0.36 & 0.90 & 0.00 & 0.10 & 0.35 &  0.95  \\
  50 & 100 & 5 &20.00 & 0.69 & 0.10 & 0.17 & 6.39 & 4.83 & 0.02 & 0.83 & 0.00 & 0.17 & 0.04 &  0.94  \\
  50 & 200 & 5 & 5.00 & 0.81 & 0.11 & 0.02 & 5.42 & 4.99 & 0.59 & 0.99 & 0.00 & 0.01 & 0.57 &  0.95  \\
  50 & 200 & 5 &10.00 & 0.78 & 0.11 & 0.06 & 5.71 & 4.99 & 0.36 & 0.99 & 0.00 & 0.01 & 0.35 &  0.95  \\
  50 & 200 & 5 &20.00 & 0.69 & 0.11 & 0.17 & 6.58 & 4.98 & 0.03 & 0.98 & 0.00 & 0.02 & 0.04 &  0.94  \\
  50 & 400 & 5 & 5.00 & 0.81 & 0.11 & 0.02 & 5.54 & 5.00 & 0.49 & 1.00 & 0.00 & 0.00 & 0.47 &  0.95  \\
  50 & 400 & 5 &10.00 & 0.78 & 0.11 & 0.06 & 5.71 & 5.00 & 0.36 & 1.00 & 0.00 & 0.00 & 0.35 &  0.95  \\
  50 & 400 & 5 &20.00 & 0.69 & 0.11 & 0.17 & 6.66 & 5.00 & 0.03 & 1.00 & 0.00 & 0.00 & 0.04 &  0.94  \\

\hline\hline
\end{tabular}
\end{center}
Notes: $X_{it}=F_t^{0^\prime}\Lambda_i^0+e_{it}+s_{it}$, $e_{it}\sim (0,1)$, $s_{it}\sim (0,\omega^2)$. $F_t$ is $r\times 1$, $r=5$. Let $C^0=F^0\Lambda^{0\prime}=U_rD_rV_r^\prime$. Then  $r^*=\sum_{j=1}^r 1(\frac{d_i^2}{\sum_{k=1}^r d_i^2}>\gamma)$ with $\gamma=0.05$,  $C^r=\frac{\var(C^0)}{\var(X)}$, $C_r=\frac{\var(F_r\Lambda_r^\prime)}{\var(X)}$, $c(S)=\frac{\var(S)}{\var(X)}$. The column labeled `spanning' is the $R^2$ from a regression of the smallest factor on $U_r$.

\newpage

% mc2.m produces mc2.out
\begin{center}
Table 2: DGP 2
\vspace{0.1in}

\begin{tabular}{llll|ll|l|ll|ll|ll|ll}
 \multicolumn{4}{c}{parameters} & \multicolumn{2}{c}{signal} & noise & \multicolumn{2}{c}{mean} &
 \multicolumn{2}{c}{prob. $=r$} & \multicolumn{2}{c}{prob. $=r^*$} & \multicolumn{2}{c}{spanning}\\
 \hline
 $N$ & $T$ & $r^*$ & $\omega$ & $C^r$  & $C_r$ & $c(S)$ & $\hat r$ & $\bar r$ & $\hat r$ &
 $\bar r$ & $\hat r$ & $\bar r$ & $\hat R^2$ & $\bar R^2$\\ \hline
 100 & 100 & 3 & 1.00 & 0.67 & 0.02 & 0.00 & 3.94 & 3.00 & 0.00 & 0.00 & 0.06 & 1.00 & 0.07 &  0.95  \\
 100 & 100 & 3 & 0.75 & 0.67 & 0.01 & 0.00 & 3.95 & 3.00 & 0.00 & 0.00 & 0.05 & 1.00 & 0.05 &  0.95  \\
 100 & 100 & 3 & 0.50 & 0.67 & 0.01 & 0.00 & 3.97 & 3.00 & 0.00 & 0.00 & 0.03 & 1.00 & 0.04 &  0.95  \\
 100 & 200 & 3 & 1.00 & 0.67 & 0.02 & 0.00 & 4.01 & 3.00 & 0.01 & 0.00 & 0.00 & 1.00 & 0.00 &  0.95  \\
 100 & 200 & 3 & 0.75 & 0.67 & 0.01 & 0.00 & 4.00 & 3.00 & 0.00 & 0.00 & 0.00 & 1.00 & 0.00 &  0.95  \\
 100 & 200 & 3 & 0.50 & 0.67 & 0.01 & 0.00 & 4.00 & 3.00 & 0.00 & 0.00 & 0.00 & 1.00 & 0.00 &  0.95  \\
 100 & 400 & 3 & 1.00 & 0.67 & 0.02 & 0.00 & 4.26 & 3.00 & 0.26 & 0.00 & 0.00 & 1.00 & 0.00 &  0.95  \\
 100 & 400 & 3 & 0.75 & 0.67 & 0.01 & 0.00 & 4.00 & 3.00 & 0.00 & 0.00 & 0.00 & 1.00 & 0.00 &  0.95  \\
 100 & 400 & 3 & 0.50 & 0.67 & 0.01 & 0.00 & 4.00 & 3.00 & 0.00 & 0.00 & 0.00 & 1.00 & 0.00 &  0.95  \\
  50 & 100 & 3 & 1.00 & 0.67 & 0.02 & 0.00 & 3.55 & 2.57 & 0.00 & 0.00 & 0.45 & 0.57 & 0.41 &  0.93  \\
  50 & 100 & 3 & 0.75 & 0.67 & 0.01 & 0.00 & 3.60 & 2.62 & 0.00 & 0.00 & 0.40 & 0.62 & 0.37 &  0.93  \\
  50 & 100 & 3 & 0.50 & 0.67 & 0.01 & 0.00 & 3.64 & 2.66 & 0.00 & 0.00 & 0.36 & 0.66 & 0.33 &  0.93  \\
  50 & 200 & 3 & 1.00 & 0.67 & 0.02 & 0.00 & 3.95 & 2.97 & 0.00 & 0.00 & 0.06 & 0.97 & 0.06 &  0.91  \\
  50 & 200 & 3 & 0.75 & 0.67 & 0.01 & 0.00 & 3.96 & 2.98 & 0.00 & 0.00 & 0.04 & 0.98 & 0.04 &  0.91  \\
  50 & 200 & 3 & 0.50 & 0.67 & 0.01 & 0.00 & 3.97 & 2.98 & 0.00 & 0.00 & 0.03 & 0.98 & 0.04 &  0.91  \\
  50 & 400 & 3 & 1.00 & 0.67 & 0.02 & 0.00 & 4.00 & 3.00 & 0.00 & 0.00 & 0.01 & 1.00 & 0.01 &  0.91  \\
  50 & 400 & 3 & 0.75 & 0.67 & 0.01 & 0.00 & 4.00 & 3.00 & 0.00 & 0.00 & 0.01 & 1.00 & 0.01 &  0.91  \\
  50 & 400 & 3 & 0.50 & 0.67 & 0.01 & 0.00 & 4.00 & 3.00 & 0.00 & 0.00 & 0.00 & 1.00 & 0.01 &  0.91  \\
\hline\hline

 100 & 100 & 3 & 1.00 & 0.60 & 0.02 & 0.11 & 4.81 & 2.93 & 0.81 & 0.00 & 0.00 & 0.93 & 0.01 &  0.93  \\
 100 & 100 & 3 & 0.75 & 0.59 & 0.01 & 0.11 & 4.84 & 2.95 & 0.84 & 0.00 & 0.00 & 0.95 & 0.01 &  0.93  \\
 100 & 100 & 3 & 0.50 & 0.59 & 0.01 & 0.11 & 4.86 & 2.96 & 0.86 & 0.00 & 0.00 & 0.96 & 0.01 &  0.93  \\
 100 & 200 & 3 & 1.00 & 0.60 & 0.02 & 0.11 & 5.01 & 3.00 & 0.99 & 0.00 & 0.00 & 1.00 & 0.01 &  0.93  \\
 100 & 200 & 3 & 0.75 & 0.59 & 0.01 & 0.11 & 5.00 & 3.01 & 1.00 & 0.00 & 0.00 & 0.99 & 0.01 &  0.93  \\
 100 & 200 & 3 & 0.50 & 0.59 & 0.01 & 0.11 & 5.00 & 3.01 & 1.00 & 0.00 & 0.00 & 0.99 & 0.01 &  0.93  \\
 100 & 400 & 3 & 1.00 & 0.60 & 0.02 & 0.11 & 5.21 & 3.10 & 0.80 & 0.00 & 0.00 & 0.90 & 0.00 &  0.84  \\
 100 & 400 & 3 & 0.75 & 0.59 & 0.01 & 0.11 & 5.00 & 3.12 & 1.00 & 0.00 & 0.00 & 0.88 & 0.00 &  0.83  \\
 100 & 400 & 3 & 0.50 & 0.59 & 0.01 & 0.11 & 5.00 & 3.13 & 1.00 & 0.00 & 0.00 & 0.87 & 0.00 &  0.82  \\
  50 & 100 & 3 & 1.00 & 0.60 & 0.02 & 0.11 & 4.18 & 2.27 & 0.34 & 0.00 & 0.16 & 0.27 & 0.16 &  0.94  \\
  50 & 100 & 3 & 0.75 & 0.59 & 0.01 & 0.11 & 4.23 & 2.31 & 0.38 & 0.00 & 0.15 & 0.31 & 0.15 &  0.93  \\
  50 & 100 & 3 & 0.50 & 0.59 & 0.01 & 0.11 & 4.28 & 2.34 & 0.41 & 0.00 & 0.13 & 0.34 & 0.14 &  0.93  \\
  50 & 200 & 3 & 1.00 & 0.60 & 0.02 & 0.11 & 4.82 & 2.83 & 0.82 & 0.00 & 0.01 & 0.83 & 0.02 &  0.89  \\
  50 & 200 & 3 & 0.75 & 0.59 & 0.01 & 0.11 & 4.84 & 2.86 & 0.84 & 0.00 & 0.01 & 0.86 & 0.02 &  0.89  \\
  50 & 200 & 3 & 0.50 & 0.59 & 0.01 & 0.11 & 4.86 & 2.87 & 0.86 & 0.00 & 0.01 & 0.87 & 0.02 &  0.89  \\
  50 & 400 & 3 & 1.00 & 0.60 & 0.02 & 0.11 & 4.97 & 2.97 & 0.96 & 0.00 & 0.00 & 0.97 & 0.01 &  0.89  \\
  50 & 400 & 3 & 0.75 & 0.59 & 0.01 & 0.11 & 4.96 & 2.98 & 0.97 & 0.00 & 0.00 & 0.98 & 0.01 &  0.89  \\
  50 & 400 & 3 & 0.50 & 0.59 & 0.01 & 0.11 & 4.97 & 2.98 & 0.97 & 0.00 & 0.00 & 0.98 & 0.01 &  0.89  \\

  \hline\hline
\end{tabular}
\end{center}
Notes: $X=C^0+e+s$, $e_{it}\sim (0,1)$, $s_{it}\sim (0,\omega^2)$,   $C^0=U_rD_rV_r^\prime$, $D_r=[1, .8, .5, .3, .2]$. Then  $r^*=\sum_{j=1}^r 1(\frac{d_i^2}{\sum_{k=1}^r d_i^2} >\gamma)$ with $\gamma=0.05$,  $C^r=\frac{\var(C^0)}{\var(X)}$, $C_r=\frac{\var(F_r\Lambda_r^\prime)}{\var(X)}$, $c(S)=\frac{\var(S)}{\var(X)}$. The column labeled `spanning' is the $R^2$ from a regression of the smallest factor on $U_r$.

\newpage

\section*{Appendix A}

\subsection*{Linear Restrictions in \textsc{matlab}}
The factor model is $ X_{it}= F_{1t} \Lambda_{i1} + F_{2t}  \Lambda_{i2} + F_{3t} \Lambda_{i3} + e_{it}$.
Consider the restrictions (i)  (i) $\Lambda_{12}=0$, (ii)) $\Lambda_{13} = 0$,
               (iii)  (v) $\Lambda_{21}=\Lambda_{31}.$
Given unrestricted estimates of the factor loadngs \textsc{Lhatu}, the following returns the restricted loadings \textsc{Lhatrv}.

{\tiny
\begin{verbatim}
R.cmat{1}=zeros(N,r); R.cmat{1}(1,2)=1.0; R.phi{1}=0.0;
R.cmat{2}=zeros(N,r); R.cmat{2}(1,3)=1.0; R.phi{2}=0.0;
R.cmat{3}=zeros(N,r); R.cmat{3}(2,1)=1.0; R.cmat{3}(3,1)=-1;R.phi{3}=0.0;

Rvec=[];
phivec=[];
for j=1:length(R.cmat);
	Rvec=[Rvec vec(R.cmat{j})];
	phivec=[phivec; R.phi{j}];
end;
Rvec=Rvec';
dum1=kron(inv(Fhat'*Fhat),eye(N));
dum2= Rvec*dum1*Rvec';
adj=dum1*Rvec'*inv(dum2)*(Rvec*vec(Lhatu)-phivec);
Lhatrv=vec(Lhatu)-adj;

% one more iteration
L_u=Lhatu;
F_u=Fhat;
it=1; maxit=100; done=0;
while done==0 & it<=maxit;
    dum1=kron(inv(F_u'*F_u),eye(N));
    dum2= Rvec*dum1*Rvec';
    adj=dum1*Rvec'*inv(dum2)*(Rvec*vec(L_u)-phivec);
    Lamrv=vec(L_u)-adj;
    L_r=reshape(Lamrv,N,r);
    F_r=X*L_r*inv(L_r'*L_r);
    err1=norm(F_r'*F_r-F_u'*F_u,'fro');
    err2=norm(L_r'*L_r-L_u'*L_u,'fro');
    if err1+err2> 1e-8;
        F_u=F_r;
        L_u=L_r;
        disp(sprintf('%d %f %f ',it,err1,err2));
        it=it+1;
    else;
       disp(sprintf('Converged: %d %f %f ',it,err1,err2));
        done=1;
    end;
end;
disp('Converged ');
disp('estimates: unrestricted             restricted');
disp([Lhatu Lhatr]);
disp(['Restricted LL      FF']);
disp([L_r'*L_r F_r'*F_r]);
for i=1:r;
    c1=corr(Fhat(:,i),F_r(:,i));
    c2=corr(Lhatu(:,i),L_r(:,i));
    mymprint([c1 c2  ]);
end;

\end{verbatim}
}
{\small
\[
\begin{pmatrix*}[r]
 4.70& -1.13&  0.89\\
 1.21&  0.77&  2.41\\
-3.67&  0.05&  1.73\\
-1.27& -3.71&  0.45\\
 0.16& -0.81& -0.93
 \end{pmatrix*}
 \quad
 \begin{pmatrix*}[r]
  4.70&  0.00&  0.00\\
-1.23&  0.77&  2.41\\
-1.23&  0.05&  1.73\\
-1.27& -3.71&  0.45\\
 0.16& -0.81& -0.93
 \end{pmatrix*}
 \]
}
\newpage

\section*{Appendix B}

\paragraph{Proof of Lemma \ref{lem:HNT}.}

Proof of part (i).
\[ \tilde H_{NT}= (\Lp\Lambda^0/N)(\Fp\tilde F/T) D_r^{-2} \]
By the fact that $D_r^2$ is the matrix of eigenvalues of $\frac{XX\pr}{NT}$ associated with the eigenvectors
$\tilde F$ (and noting the normalization $\tilde F'\tilde F=T I_r$, we have
$ \tilde F\pr(\frac{XX\pr}{NT})\tilde F  = T D_r^2 $.
Substituting   $X=F^0\Lp + e$ into the above, we have
\[ D_r^2 =(\tilde F\pr F^0/T)(\Lp \Lambda^0/N) (\Fp \tilde F/T)  +  \frac 1 T \tilde F\pr ee\pr\tilde F/(NT).  \]
Since the second term is $o_p(1)$, we can substitute
$ D_r^{-2} =(\Fp \tilde F/T)^{-1}(\Lp \Lambda^0/N)^{-1}(\tilde F\pr F^0/T)^{-1} + o_p(1) $
into $\tilde H_{NT}$ to give
\[ \tilde  H_{NT} = (\tilde F\pr F^0/T)^{-1} +o_p(1). \]
Denote
\[ \tilde H_{1,NT}=(\Lp\Lambda^0/N)(\tilde \Lambda\pr \Lambda^0/N)^{-1} \]
Left and right multiplying $X=F^0\Lp + e$ by $\tilde F\pr$  and $\Lambda^0$ respectively, dividing  by $NT$, and using $\tilde \Lambda =\tilde F\pr X/T$, we obtain
\[ \frac {\tilde \Lambda\pr\Lambda^0} N =\frac {\tilde F\pr F^0} T \frac{\Lp \Lambda^0} N  + o_p(1). \]
Substituting
$ \Big(\frac {\tilde \Lambda\pr\Lambda^0} N\Big)^{-1} =\Big(\frac{\Lp \Lambda^0} N\Big)^{-1}\Big(\frac {\tilde F\pr F^0} T \Big)^{-1} +o_p(1)
$
into $\tilde H_{1,NT}$, we obtain
\[  \tilde H_{1,NT}= \Big(\frac {\tilde F\pr F^0} T \Big)^{-1} +o_p(1) .\]
Thus $\tilde H_{NT}$ and $\tilde H_{1,NT}$ have the same asymptotic expression. This proves part (i).

Proof of part (ii). The proof of part (i) shows that
\[ (\Lp\Lambda^0/N)(\tilde \Lambda\pr \Lambda^0/N)^{-1} =(\tilde F' F^0/T)^{-1} + o_p(1) \]
Take transpose and inverse, we have
\[ F^{0\prime}\tilde F/T = (\Lp\Lambda^0/N)^{-1} (\Lp\tilde \Lambda/N) + o_p(1) \]
Substitute this expression into the original definition of $\tilde H_{NT}$, we have
\[ \tilde H_{NT} =(\Lp\tilde \Lambda/N) D_r^{-2}  + o_p(1) \]
Now left multiply the equation $X=F^0 \Lp +e $ by $\Fp$ and right multiply it by $\tilde \Lambda$, divide by $NT$, we obtain
\[ \Fp X \tilde \Lambda/(NT) =(\Fp F^0/T) (\Lp \Lapca/N)  + \Fp e \Lapca/(NT) \]
But
\[ X\Lapca =X\Lapca (\Lapca'\Lapca)^{-1} (\Lapca'\Lapca)=\tilde F(\Lapca'\Lapca)=\tilde F D_r^2 N \]
Thus we have
\[ (\Fp \tilde F/T)D_r^2  =(\Fp F^0/T) (\Lp \Lapca/N) +o_p(1) \]
Equivalently,
\[ (\Fp F^0/T)^{-1}(\Fp \tilde F/T) = (\Lp \Lapca/N) D_r^{-2} +o_p(1) \]
But the left hand side is equal to $\tilde H_{NT}+o_p(1)$. This completes the proof of (ii).

Analogously, $\hat H_{1,NT}=\hat H_{NT}+o_p(1)$ and  $\hat H_{2,NT}=\hat H_{NT}+o_p(1)$.
Consider the first claim.
From $\hat \Lambda =\tilde \Lambda D_r^{-1/2}$, we have
\[ (\Lambda^{0\prime}\Lambda^0)(\Lpca^{\prime}   \Lambda^0)^{-1}=
(\Lambda^{0\prime}\Lambda^0)(\Lapca^{\prime}   \Lambda^0)^{-1} D_r^{1/2} =\tilde H_{NT} D_r^{1/2} +o_p(1)
=\hat H_{NT} + o_p(1) \]
the second equality uses  Lemma \ref{lem:HNT}(i), and the last equality uses the definition of $\hat H_{NT}$.
The proof of the second claim is similar by using $\hat F =\tilde F D_r^{1/2}$.

\paragraph{Proof of (\ref{bar-Lambdai}).}
\begin{eqnarray*}  \bar \Lambda_i &=  & \Delta_{NT} \hat \Lambda_i  \\
&=& \Delta_{NT} ( \hat \Lambda_i -\hat H_{NT}^{-1}\Lambda_i^0+ \hat H_{NT}^{-1}\Lambda_i^0) \\
&=& \Delta_{NT} ( \hat \Lambda_i -\hat H_{NT}^{-1}\Lambda_i^0) + \Delta_{NT} \hat H_{NT}^{-1}\Lambda_i^0\\
&=&  \Delta_{NT} ( \hat \Lambda_i -\hat H_{NT}^{-1}\Lambda_i^0) +\Delta_{NT}^2 (\bar H_{NT})^{-1}\Lambda_i^0
\end{eqnarray*}
%\begin{eqnarray*}  \bar \Lambda_i &= &(D^\gamma_r)^{1/2} D_r^{-1} \tilde \Lambda_i \\
%&=&  (D^\gamma_r)^{1/2} D_r^{-1} [\tilde \Lambda_i- \tilde H_{NT}^{-1}\Lambda_i^0+ \tilde H_{NT}^{-1}\Lambda_i^0] %\\ &=& (D^\gamma_r)^{1/2} D_r^{-1} [\tilde \Lambda_i- \tilde H_{NT}^{-1}\Lambda_i^0] +
%(D^\gamma_r)^{1/2} D_r^{-1}\tilde H_{NT}^{-1} \Lambda_i^0 \\
%&=& (D^\gamma_r)^{1/2} D_r^{-1} [\tilde \Lambda_i- \tilde H_{NT}^{-1}\Lambda_i^0] +
%(D^\gamma_r)^{1/2} D_r^{-1} (D^\gamma_r)^{1/2} (\bar H_{NT}^\gamma)^{-1} \Lambda_i^0
%\end{eqnarray*}
Moving the second term to the left hand side, we obtain (\ref{bar-Lambdai}).

\paragraph{Proof of (\ref{common-distribution}).}
For notational simplicity, write $\bar H$ for $\bar H_{NT}$,  and $\bar G$ for $\bar G_{NT}$.
\begin{eqnarray*}  \bar C_{it}-\Fp_t \bar H  \, \bar G \Lambda_i^0 & = & \\
 &= & (\bar F_t-\bar H'F_t^0)'\bar \Lambda_i + \Fp_t\bar H (\bar\Lambda_i -\bar G \Lambda_i^0)\\
 &= & \sqrt{T} (\bar F_t-\bar H'F_t^0)'\, (\bar \Lambda_i/T^{1/2})  +  (\Fp_t \bar H/N^{1/2})\,  \sqrt{N} (\bar\Lambda_i  -\bar G \Lambda_i^0)
 \end{eqnarray*}
 From Proposition \ref{prop:prop2},
 \[\sqrt{T} (\bar F_t-\bar H'F_t^0) \dconv N(0, Avar(\bar F_t)), \quad \sqrt{N} (\bar\Lambda_i  -\bar G \Lambda_i^0)\dconv N(0,Avar(\bar \Lambda_i)) \]
 The two asymptotic distributions are independent because the first involves sum of random variables over the cross sections for period $t$, and the second distribution involves random variables over all time periods for individual  $i$. Let
 \[ \bar A_{NT}= \frac 1 T \bar \Lambda_i' Avar(\bar F_t) \bar \Lambda_i + \frac 1 N \bar F_t' Avar(\bar \Lambda_i) \bar F_t \]
 If we replace  $Avar(\bar F_t)$ and $ Avar(\bar \Lambda_i)$ by their estimated versions, then $\bar A_{NT}$ is the estimated variance of $\bar C_{it}-\Fp_t \bar H  \, \bar G \Lambda_i^0$.
 Using argument  similar to Bai (2003), we have
 \[ (\bar A_{NT})^{-1/2}(\bar C_{it} -\Fp_t \bar H \bar G \Lambda_i^0)  \dconv N(0,1)  \]
 Since
 \[ \bar H \, \bar G = \bar H \Delta_{NT}^2 \bar H^{-1} =\bar H (I_r -\gamma D_r^{-1}) \bar H^{-1}= I_r-\gamma \bar H D_r^{-1} \bar H^{-1}\]
 This gives
 \[ \Fp_t \bar H \, \bar G \Lambda_i^0 =\Fp_t\Lambda_i^0-\gamma \Fp_t \bar H D_r^{-1} \bar H^{-1} \Lambda_i^0 =C_{it}^0-\gamma \Fp_t \bar H D_r^{-1} \bar H^{-1} \Lambda_i^0 \]
 Thus
 \[ (\bar A_{NT})^{-1/2}(\bar C_{it}-C_{it}^0-bias ) \dconv N(0,1) \]
 where $bias = \gamma \Fp_t \bar H D_r^{-1} \bar H^{-1} \Lambda_i^0 $.

\paragraph{Proof of Proposition \ref{prop:prop3}.} The estimator by direction calculations is given by
\begin{eqnarray*}
\label{RR-estimator}
 \vec(\Lcpca)&=& \vec(\Lridge) -\Delta(\gamma,\phi)
 \end{eqnarray*}
 where
 $\Delta(\gamma,\phi)=
 ((F^{+}\otimes I_N)^\prime (F^+\otimes I_N))^{-1} R' \Big[ R\Big((F^+\otimes I_N)^\prime (F^+\otimes I_N)\Big)^{-1}R'\Big]^{-1} \Big( R \, \vec(\Lridge)-\phi\Big).$
But
\begin{eqnarray*}
 \bigg((F^{ +}\otimes I_N)^\prime (F^+\otimes I_N)\bigg)^{-1} R'&=&   \Big((F'F+\gamma  I_r)^{-1}\otimes I_N\Big)  R' \\
  R\bigg[(F^+\otimes I_N)^\prime (F^+\otimes I_N)\bigg]^{-1} R'&=&  R\Big[(F'F+\gamma  I_r)^{-1}\otimes I_N \Big] R'.
 \end{eqnarray*}
Hence
\[ \Delta(\gamma,\phi)= [(F'F+\gamma  I_r)^{-1}\otimes I_N] R'
 \cdot  \Big[ R[(F'F+\gamma  I_r)^{-1}\otimes I_N] R'\Big]^{-1}\Big( R \, \vec(\Lridge)-\phi\Big).\]

\paragraph{Proof of (\ref{generalize-RPCA-F}) and (\ref{generalize-RPCA-L}).}
With $\tau=0$, the objective function becomes
\[ \|Z-F\Lambda'\|_F^2 +\gamma_1 \|F\|_F^2 +\gamma_2 \|\Lambda\|_F^2  \]
Consider a change of  variables $F=(\gamma_2/\gamma_1)^{1/4}\ddot F$ and $\Lambda=(\gamma_1/\gamma_2)^{1/4}\ddot \Lambda$. Then the objective function can be rewritten as
\[ \|Z-\ddot F \ddot \Lambda'\|_F^2 +\sqrt{\gamma_1\gamma_2} \|\ddot F\|_F^2 +\sqrt{\gamma_1\gamma_2} \|\ddot \Lambda\|_F^2  \]
This is an objective function with equal weights, we know the optimal solution is
\[ \ddot F =U_r [(D_r-\sqrt{\gamma_1\gamma_2} I_r)_+]^{1/2}, \quad \ddot \Lambda =V_r [(D_r-\sqrt{\gamma_1\gamma_2} I_r)_+]^{1/2}  \]
In terms of the original variables, the optimal solution is
\[ F=(\gamma_2/\gamma_1)^{1/4} U_r [(D_r-\sqrt{\gamma_1\gamma_2} I_r)_+]^{1/2}, \quad  \Lambda =(\gamma_2/\gamma_1)^{1/4}V_r [(D_r-\sqrt{\gamma_1\gamma_2} I_r)_+]^{1/2}  \]
This gives  (\ref{generalize-RPCA-F}) and (\ref{generalize-RPCA-L}).

 \newpage
 \baselineskip=12.0pt
\bibliography{macro,metrics,kn,factor,ll,metrics2,bigdata}

\end{document}